\documentclass[lineno]{article}
\usepackage{arxiv}
\usepackage{graphicx}
\usepackage{epstopdf, epsfig}
\usepackage{amssymb,amsmath}
\usepackage{tikz}
\usepackage{graphicx}
\usepackage{dsfont}
\usepackage{accents}
\usepackage{pgfplots}
\usepackage{hyperref}
\usepackage{labelfig,epsf}
\usepackage{caption, subcaption}
\usepackage[export]{adjustbox}
\usepackage{float}
\usepackage{orcidlink}
\usepackage{authblk}
\usepackage{blindtext}
\pgfplotsset{compat=1.18}
\newlength{\dhatheight}

\usepackage{newtxtext}
\usepackage{newtxmath}
\usepackage{natbib}
\usepackage{hyperref}
\hypersetup{
	colorlinks = true,
	urlcolor   = blue,
	citecolor  = black,
}

\newcommand{\RomanNumeralCaps}[1]
\linenumbers

\renewcommand{\exp}{\mathrm{e}}
\newcommand{\ci}{\mathrm{i}}
\newcommand{\im}{\mathrm{i}}
\newcommand{\wrt}{\mathrm{d}}

\newcommand{\doublehat}[1]{%
	\settoheight{\dhatheight}{\ensuremath{\hat{#1}}}%
	\addtolength{\dhatheight}{-0.25ex}%
	\hat{\vphantom{\rule{1pt}{\dhatheight}}%
		\smash{\hat{#1}}}}

\begin{document}
	\title{Attenuation of long waves through regions of irregular floating ice and bathymetry}
	
	\author{\textbf{Lloyd Dafydd}\thanks{Corresponding Author: lloyd.dafydd@bristol.ac.uk}}
	\author{\textbf{Richard Porter}}
	\affil{School of Mathematics, Woodland Road, University of Bristol, Bristol, BS8 1UG, UK}
	\maketitle
	
	\begin{abstract}
		Existing theoretical results for attenuation of surface waves
		propagating on water of 
		random fluctuating depth are shown to over predict the rate of decay
		due to the way in which ensemble averaging is performed. 
		A revised approach is presented
		which corrects this and is shown to conserve energy. New theoretical 
		predictions are supported by numerical results which use averaging of
		simulations of wave scattering over finite sections of random bathymetry for which
		transfer matrix eigenvalues are used to accurately measure decay.
		The model of wave propagation used 
		in this paper is derived from a linearised long wavelength assumption whereby 
		depth averaging leads to time harmonic waves being represented as solutions to
		a simple ordinary differential equation. In this paper it is shown how
		this can be adapted to incorporate a model of a continuous covering of
		the surface by fragmented floating ice. 
		Attenuation of waves through broken ice of random thickness is then 
		analysed in a similar manner as bed variations previously. General features of the predicted attenuation are discussed in relation to existing theoretical models for attenuation due to multiple scattering through random ice environments and to field data, particularly in the models ability to capture a ``rollover effect'' at higher frequencies.
	\end{abstract}
	
	\textbf{Keywords}
		Wave scattering;
		Shallow water flows; 
		Sea ice.

	\section{Introduction}\label{Introduction}
	
	It is well known that waves become attenuated as they propagate through an inhomogeneous 
	disordered medium that has randomly varying properties. The term ``localisation'' is used to describe this phenomenon since the waves are localised in space. Localisation is 
	recognised as a multiple scattering effect caused by incoherent 
	reflections from within the disordered medium and is an energy conserving
	process; that is, attenuation is not a feature of natural physical dissipative effects. 
	
	The pioneering work of 
	\citet{Anderson_1958} which first described 
	localisation in quantum systems has since
	been applied to many other physical systems supporting 
	wave motion. Amongst these, considerable attention has been paid to the 
	propagation of water waves over randomly-varying
	bathymetry and this is the main initial focus of this paper.
	Early work in this area considered the randomness be
	manifested by rectangular steps in the bed. Following the experiments 
	of \citet{Belzons_Guazzelli_Parodi_1988}, \citet{Devillard_Dunlop_Souillard_1988} used both shallow water and wide-spacing analogous full linear potential
	theory to consider the effect of random stepped bathymetry on wave
	propagation. Their numerical results supported 
	an asymptotic theory based on a long wavelength assumption that
	attenuation (the spatial rate of decay and the reciprocal
	of localisation length) is proportional to the square of the
	wave frequency. For longer waves, their numerical results based on
	shallow-water theory diverged, unsurprisingly, from the asymptotic 
	long wavelength theory and from numerical simulations based on full 
	potential theory, and indicated that attenuation tended to a constant for high frequencies.
	Full linear potential theory suggested otherwise: that 
	attenuation becomes 
	exponentially weak as wavelengths tend towards the short wavelength regime and this was
	explained as being associated with the exponential decay of wave
	energy throughout the fluid depth.
	
	Other work on random beds worthy of note include a series of
	papers by Nachbin and co-authors (see \citet{nach_pap92a,nach_pap92b,nach95}).
	Much of the work on waves over random beds have supported the
	findings outlined above. Within a linearised setting  
	\citet[Section~7.4]{Mei_Stiassnie_P._2005} applies a 
	multiple-scales method (based on the work of 
	\citet{Kawahara_Yoshimura_Nakagawa_Ohsaka_1976}) for non-shallow potential flow 
	and reaches similar conclusions. The calculation results in
	an explicit formula for the attenuation rate which is linked
	to the assumed statistical properties of the bed (now assumed to be defined by a smoothly varying function), as well as
	wavelength and the mean water depth. Around the same time,
	a number of papers (see \citet{Pihl_Mei_Hancock_2002}, \citet{Grataloup_Mei_2003}, 
	\citet{Mei_Li_2004}) applied similar multiple-scales analysis to various nonlinear descriptions of wave propagation. In particular 
	\citet{Mei_Li_2004} and \citet{Grataloup_Mei_2003} considered weakly nonlinear long wavelength theories (Boussinesq approximations). 
	The analytically-derived formulae for wave attenuation
	differed in that it predicted attenuation increasing like
	the frequency squared across all frequencies. Thus, there is no levelling off in the attenuation as described by \citet{Devillard_Dunlop_Souillard_1988}
	nor exponential decay as predicted by full linear potential theory.
	
	More recently, 
	\citet{Bennetts_Peter_Chung_2014} returned
	to the problem of linear full potential theory and performed a 
	series of careful numerical simulations, over stepped beds, which they compared to the theory described by
	\citet[Section~7.4]{Mei_Stiassnie_P._2005}. They estimated the attenuation of individual waves, averaged over different realisations of random bathymetry and showed attenuation is significantly weaker than predicted by the theory. They correctly conclude that the ensemble averaging process used in the multiple-scales analysis contributes to an over-prediction of the decay of wave energy due to phase cancellation of propagating waves.
	\citet{Bennetts_Peter_Chung_2014} also attempted to correct for the failings of the existing modelling by including both left- and right-going waves in the leading order solution and by assuming a dependence on the random variables (i.e. stochastic) in the leading order solution, as opposed to making the usual assumption that it is deterministic.
	
	In this paper we revisit the problem of scattering by random bathymetry using a long wavelength/shallow water model which reduces the scattering process to solving an ordinary differential equation (ODE) that includes a coefficient of a random variable with given statistical properties (see Section \ref{Section 3}). In particular the random variations in height are considered small compared to the depth. Our analysis (Section \ref{Section 4}) is different to previous approaches. First, we assume the randomness occupies a semi-infinite region and define the problem in terms of an incident wave which has the effect of introducing an energy budget. Like \citet{Bennetts_Peter_Chung_2014} we include left- and right-propagating waves, but we assume the leading order solution is deterministic. Like \citet[Section~7.4]{Mei_Stiassnie_P._2005} (and others) we adopt a multiple-scales approach, but note that the ensemble averaging which determines the attenuation requires careful consideration to remove phase cancellations which are not associated with multiple scattering. In making this correction we also show that energy is conserved. 
	
	Theory is compared to numerical simulations which are described in 
	Section \ref{Section 5} of the paper. In Section \ref{Section 6} we use an extension of the model (derived in the Appendix) which allows for the surface of the water to be entirely covered by fragmented ice of variable thickness. The ODE that results differs from the variable bathymetry case only in the definition of three scaling coefficients and a dispersion relation; theory and numerical results are compared in Section \ref{Section 7} of the paper. 
	
	There are a number of existing studies in the literature that have 
	explored the relationship between attenuation as a result of multiple 
	scattering through randomness in ice. Only a few are three dimensional
	(e.g. \citet{bennettsetal2010} and \citet{Montiel_Squire_Bennetts_2016}) and most make the same two-dimensional 
	simplification made here. Others such as \citet{MOSIG2019153} have derived one-dimensional models in the form of a transport equation derived from the work of \citet{RYZHIK1996327} investigating elastic waves in random media. Attenuation due to changes in the thickness
	of ice were considered by \citet{kohoutmeylan} who
	represent ice floes as a series of thin elastic plates with free edges
	floating in the surface with zero (non-Archimedian) draught. Additional dissipation models related to dependence on ice thickness were considered by \citet{YU2022103582}, who derived a non-linear model dependent on ice thickness, \citet{jmse10101472}, who considered Reynolds stress in a two-layer fluid system, and \citet{SUTHERLAND2019111}, who used dimensional analysis under the assumption of their being some self-similarity scaling law. The floes 
	are considered sufficiently long to make a wide-spacing approximation 
	(\cite{PORTER2006425} showed this requires the length of the floes to 
	be of the order of the wavelength for this approximation to hold) and 
	averaging is performed over randomly-varying length (see \citet{WilliamsThesis})
	to avoid coherent resonant effects. Furthermore the serial 
	transmission method of \citet{wadhams_etal_1988} is used in which reflections at
	each ice edge are discarded, leading to attenuation being equated
	to accumulated transmission across multiple floes. \citet{squire2009}
	built on the work of \citet{kohoutmeylan} using data on the thickness
	of ice from a 1670km transect of the Arctic ocean. They also
	included a damping term in their plate equation following
	\citet{vaughan} whose role was intended to capture
	some natural physical dissipative effects. This approach neglects 
	an associated frequency dependence which depends on the physical
	damping process being modelled and
	its contribution to attenuation is easily seen to be proportional
	to $\omega^2$. The results claimed that multiple scattering 
	dominates at low periods and damping at higher periods.
	The method of \citet{kohoutmeylan} is extended further in
	\citet{BENNETTS20121} to include the
	effects of cracks, leads and pressure ridges. Scattering from 
	these more sophisticated features are parametrised and the
	overall attenuation from all three features are blended using
	the method of \citet{dumont}.
	
	All the models predict some attenuation which is frequency dependent
	but, without introducing a damping term of non-physical origin
	into the boundary conditions
	(see \citet{Meylan_Bennetts_Mosig_Rogers_Doble_Peter_2018} who discuss
	the ``Robinson-Palmer model''),
	no model has yet successfully replicated the field measurements;
	see discussions in \citet{montiel_etal22}, \citet{Meylan_Bennetts_Mosig_Rogers_Doble_Peter_2018}. Another feature of the field data is the onset
	of a high-frequency rollover effect in which the attenuation peaks
	and then appears to decrease as the frequency increases past a 
	critical frequency. Recently \citet{thomson_etal21} have provided
	evidence
	that the rollover effect may be a byproduct of instrument noise 
	as opposed to a physical effect.
	
	In the final part of Section \ref{Section 7} we discuss the general
	features exhibited by our model and how these relate to the models
	and the field data discussed above, taking care to note that 
	our modelling assumptions of shallow water and a continuum description
	of the broken ice cover have limitations.
	Finally, the work is summarised in Section \ref{Conclusions}.
	
	\section{Summary of the model}\label{Section 2}
	
	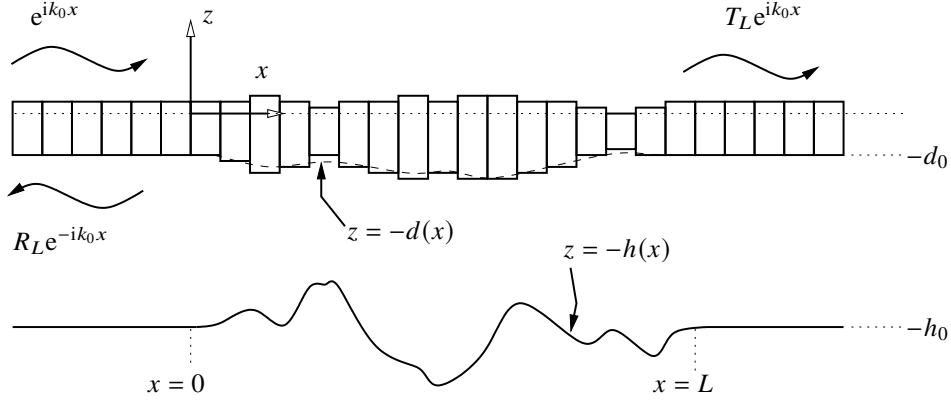
\begin{figure}
		\begin{center}
			\SetLabels
			\L (0.28*0.83) $x$ \\
			\L (0.22*0.98) $z$ \\
			\L (1.00*0.60) $-d_0$ \\
			\L (1.00*0.14) $-h_0$ \\
			\L (0.16*-0.01) $x=0$ \\
			\L (0.72*-0.01) $x=L$ \\
			\L (0.38*0.40) $z = -d(x)$ \\
			\L (0.62*0.35) $z = -h(x)$ \\
			\L (0.03*0.97) $\exp^{\ci k_0 x}$ \\
			\L (0.80*0.97) $T_L \exp^{\ci k_0 x}$ \\
			\L (0.01*0.37) $R_L\exp^{-\ci k_0 x}$ \\
			\endSetLabels
			\leavevmode
			\strut{\AffixLabels{\includegraphics[width=120mm]{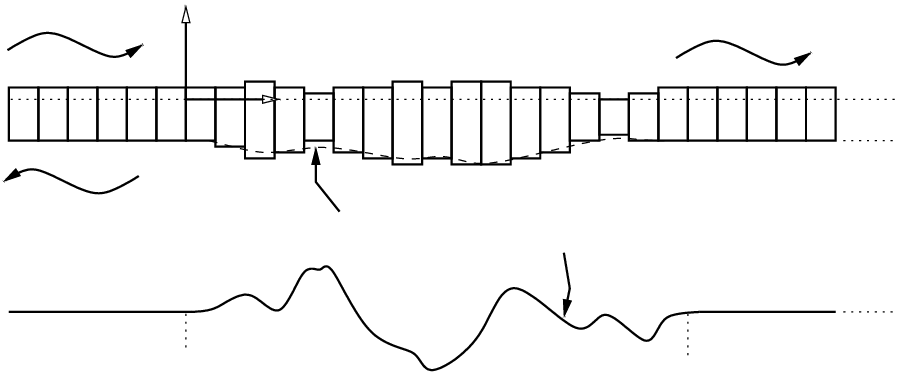}}}
			\caption{\label{fig1} Definition sketch of variable floating broken ice over a variable bed.}
		\end{center}
	\end{figure}
	
	We consider a two-dimensional scattering problem in which 
	plane-crested monochromatic waves of small amplitude propagate
	in the positive $x$-direction in $x < 0$ over fluid of 
	constant depth with a surface covered by a continuous layer of fragmented ice of 
	constant thickness. There are no physical mechanisms included in the model for
	energy dissipation such as fluid viscosity or ice-ice friction. 
	Incident wave energy is partially reflected from,
	and partially transmitted into, the region $x >0$. This is due to either
	randomly-varying bathymetry or by randomly-varying thickness of
	broken ice (both are illustrated in Fig.~\ref{fig1}) which extends over the interval $0 < x < L$ before
	returning, in $x > L$, to the same constant values found in $x < 0$.
	We are interested in monitoring the reflected and transmitted
	wave energy. 
	In Section \ref{Section 4} we set $L = \infty$ so that the randomness 
	extends indefinitely into $x > 0$. In this case all incoming wave 
	energy will be reflected and the focus is determining the attenuation
	of waves as a function of distance into $x>0$.
	
	\citet{porter_2019} developed a shallow-water (long wavelength) model for
	wave scattering over variable bathymetry with no ice cover. This
	model results from an expansion to second order in a small
	parameter representing the ratio of vertical to horizontal 
	lengthscales combined with depth averaging and is expressed by
	\begin{equation}
		(\doublehat{h}(x) \Omega'(x))' + K \Omega(x) = 0
		\label{eqn1.1}
	\end{equation}
	where $K = \omega^2/g$, $\omega$ is the angular frequency of the
	motion, $g$ represents gravitational acceleration and
	\begin{equation}
		\doublehat{h}(x) = \frac{h(x)(1-\frac13 K h(x))}{1 + \frac13 v(h) h'^2(x)}
		\label{eqn1.2}
	\end{equation}
	is defined in terms of the fluid depth $h(x)$. Here, $v(h) = 
	1 + \frac{1}{12} Kh(x)/(1-\frac13 Kh(x))$ and $v(h) \approx 1$ is a simplification 
	which will be adopted hereafter. The underlying assumptions are expressed by
	the formal constraint that $K h \ll 1$, although
	\citet{porter_2019} showed
	by comparing with exact results for reflected and transmitted
	wave energy for shoaling beds of finite length, that the model produces
	accurate predictions up to $Kh \approx 1$. 
	
	The dependent variable, $\Omega$, in (\ref{eqn1.1}) is related to the 
	time-independent wave elevation $\eta(x)$ obtained under the time-harmonic assumption $\zeta(x,t) = \Re \{ \eta(x) \exp^{-\ci \omega t} \}$ by
	\begin{equation}
		\eta(x) = \frac{-(\ci /\omega)}{\sqrt{1 - \frac13 K h(x)}}
		\left( \Omega(x) - \frac{\frac16 h h'}{1+\frac13 h'^2} \Omega'(x) \right).
		\label{eqn1.3}
	\end{equation}
	and is referred to as the ``pseudo-potential'' by \citet{TOLEDO_AGNON_2010}.
	It was shown in \citet{porter_2019} that $\Omega(x)$ and $\Omega'(x)$
	remain continuous at discontinuities in $h'(x)$. 
	
	\citet{porter_2019} highlighted the significant improvement in results away from the zero frequency limit that could be achieved when $\doublehat{h}(x) = h(x)$ is replaced by the definition in (\ref{eqn1.2}), applying in the case of the standard 
	linear shallow water equation. Thus, the modification
	in (\ref{eqn1.2}) includes, in the numerator, the effect of weak dispersion
	and, in the denominator, a geometric factor indicating a reduction
	in wave speed over sloping beds. We also remark that (\ref{eqn1.1}) can also
	be derived from a linearisation of Boussinesq equations (e.g. \citet{peregrine76}) whereby wave amplitudes are assumed sufficiently small compared to $Kh$.
	
	In the Appendix, the model developed by \citet{porter_2019} is
	extended to include the additional effect of a floating fragmented
	ice cover. Additional assumptions apply here. Ice is assumed to completely
	cover the surface of the fluid and is broken into sections which are 
	sufficiently small in horizontal extent and whose thickness 
	varies slowly enough that the submergence of the ice is represented by a continuous function, $d(x)$. Thus, the model is simulating the effect of randomness within the ice cover as rather than from incoming waves approaching the cover.
	The motion of the ice is constrained in heave (vertical) motion and the 
	expansion to second-order of the depth ratio ($\epsilon$ in the Appendix) in the modelling is needed to include the
	effect of inertia of floating ice. That is, a basic first-order linear shallow-water 
	model neglects vertical accelerations and the effect of ice cover
	at leading order is manifested only through a reduction in the depth of the fluid 
	from $h(x)$ to $h(x) - d(x)$. Thus, our second-order
	model extended to incorporate floating ice of submergence $d(x)$ is, see (\ref{eqnA.38}),
	\begin{equation}
		(\doublehat{d}(x) \Omega'(x))' + K \Omega(x) = 0,
		\label{eqn1.4}
	\end{equation}
	where $\doublehat{d}(x)$ is defined by (\ref{eqnA.39}) and the loaded surface
	elevation is related to $\Omega$ by (\ref{eqnA.40}). As before, $\Omega$ and 
	$\Omega'$ are continuous even if $d'(x)$ and/or $h'(x)$ is discontinuous.
	
	In $x < 0$ and in $x > L$ we assume $h = h_0$, $d = d_0$ are both constant.
	Then (\ref{eqn1.4}) can be solved explicitly and 
	\begin{equation}
		\Omega(x) = \exp^{\ci k_0 x} + R_L \exp^{\ci k_0 x}, \qquad
		x< 0
		\label{eqn1.7}
	\end{equation}
	\begin{equation}
		\Omega(x) = T_L \exp^{\ci k_0 x}, \qquad x > L
		\label{eqn1.8}
	\end{equation}
	where $R_L$ and $T_L$ are reflection and transmission coefficients,
	satisfying $|R_L|^2 + |T_L|^2 = 1$ (energy conservation) and
	\begin{equation}
		k_0^2 (h_0 - d_0) = \frac{K}{1 - \frac13 K (h_0 + 2 d_0)}
		\label{eqn1.9}
	\end{equation}
	defines the wavenumber, $k_0$, in terms of the frequency $\omega$.
	This shallow water dispersion relation is weakly dispersive, 
	but for sufficiently small frequencies we note that $k_0 \propto \omega$.

	\section{Description of randomness}\label{Section 3}
	
	We will consider wave propagation over a region $0 < x < L$ in which
	either the bed or the ice thickness randomly varies. We could consider
	both simultaneously varying, but for clarity consider the two
	effects separately. 
	
	We say that either
	\begin{equation}
		d = 0, \qquad h(x) = h_0(1+ \sigma r(x))
		\label{eqn2.1}
	\end{equation}
	or that
	\begin{equation}
		h = h_0, \qquad d(x) = d_0(1 + \sigma r(x))
		\label{eqn2.2}
	\end{equation}
	such that $r(x)$ is a random function with mean zero and
	unit variance. That is
	\begin{equation}
		\langle r \rangle = 0, 
		\qquad
		\langle r^2 \rangle = 1,
		\label{eqn2.3}
	\end{equation}
	implying that $\sigma$ is the RMS of the vertical variations of $h(x)$
	or $d(x)$. We ensure that the $r(0) = r'(0) = r(L) = r'(L) = 0$ so that
	the bed/ice thickness joins the constant values in $x < 0$ and $x > L$ 
	smoothly. The random function $r(x)$ also satisfies the Gaussian 
	correlation relation
	\begin{equation}
		\langle r(x) r(x') \rangle = \exp^{-|x-x'|^2/\Lambda^2}
		\label{eqn2.4}
	\end{equation}
	(other models have used an exponential correlation function, but show 
	that it produces only small differences in results). Thus, $\Lambda$ characterises the horizontal lengthscale of the random bed fluctuations.
	
	\section{Analysis of the model}\label{Section 4}
	
	In this section, we assume $L \to \infty$ so that the randomness occupies $x > 0$. 
	The main assumption that is made is that the amplitude of the randomness 
	is small, i.e. $\sigma \ll 1$. We assume $\sigma = O(\epsilon)$ and will expand up to $O(\sigma^2)$ to be consistent with the $O(\epsilon^2)$ expansion derived in the Appendix. We note that we can write (\ref{eqn1.4}) with
	(\ref{eqnA.39}), (\ref{eqnA.41}) and either (\ref{eqn2.1}) or (\ref{eqn2.2}) as
	\begin{equation}
		((1+ \sigma C_1 r(x) - \sigma^2 (C_2r^2(x)
		+ C_3 r'^2(x))) \Omega')' + k_0^2 \Omega = 0, \qquad x > 0
		\label{eqn3.1}
	\end{equation}
	where terms up to $O(\sigma^2)$ have been retained, and
	\begin{equation}
		\Omega'' + k_0^2 \Omega = 0, \qquad x < 0
		\label{eqn3.2}
	\end{equation}
	where $k_0$ is defined by (\ref{eqn1.9}).
	In (\ref{eqn3.1}), the coefficients depend on the whether the bed or the
	thickness of floating ice is represented by the random function $r(x)$.
	In the case that the bed is varying and the ice is absent, $d_0 = 0$ and
	\begin{equation}
		C_1 = \frac{1-{\textstyle \frac23} Kh_0}{1 - {\textstyle \frac13} Kh_0}, 
		\quad 
		C_2 = \frac{{\textstyle \frac13} Kh_0}{1 - {\textstyle \frac13} Kh_0}, 
		\quad 
		C_3 = {\textstyle \frac13} h_0^2
		\label{eqn3.4}
	\end{equation}
	and in the case where the ice is varying and the bed is of constant depth, $h(x) = h_0$ and
	\begin{equation}
		C_1 = \frac{-d_0(1+{\textstyle \frac13}K(h_0-4d_0))}{(h_0-d_0)(1-{\textstyle \frac13}K(h_0+2d_0))},
		\quad 
		C_2 = \frac{-{\textstyle \frac23} K d_0^2}{(h_0-d_0)(1-{\textstyle \frac13}K(h_0+2d_0))}, 
		\quad 
		C_3 = {\textstyle \frac13} d_0^2.
		\label{eqn3.4b}
	\end{equation}
	The long wave assumption on which the model is based 
	formally requires $Kd_0 < Kh_0 \ll 1$ and so we do not envisage using the 
	model close to $Kh_0 = 3$ or $K(h_0+2d_0)=3$. The solution to (\ref{eqn3.2}) is
	\begin{equation}
		\Omega(x) = \exp^{\ci k_0 x} + R_\infty \exp^{-\ci k_0 x}
		\label{eqn3.5}
	\end{equation}
	and since we anticipate decay of waves into $x \to \infty$ we
	also impose $\Omega \to 0$ as $x \to \infty$ and so we must require
	that $|R_\infty| = 1$; all incident wave energy is reflected.
	
	We make the multiple scales assumption of, for e.g., \citet{Mei_Li_2004} (but also see other references listed in the introduction) and introduce a slow variable $X = \sigma^2 x$, writing
	\begin{equation}
		\Omega(x) = \Omega_0(x,X) + \sigma \Omega_1(x,X) + \sigma^2 \Omega_2(x,X) + \ldots.
		\label{eqn3.6}
	\end{equation}
	Accordingly (\ref{eqn3.1}) becomes
	\begin{multline}
		\Bigg[ \left(\frac{\partial}{\partial x} + \sigma^2 \frac{\partial}{\partial X} \right)
		\left( \left( 1 + \sigma C_1 
		r(x) - \sigma^2 (C_2r^2(x)
		+ C_3 r'^2(x)) \right)
		\left(\frac{\partial}{\partial x} + \sigma^2 \frac{\partial}{\partial X} \right) \right)
		\\
		+ k_0^2 \Bigg] (\Omega_0 + \sigma \Omega_1 + \sigma^2 \Omega_2 +\ldots ) = 0, \qquad x > 0.
		\label{eqn3.7}
	\end{multline}
	The matching conditions at $x=0$ consist of
	\begin{equation}
		\Omega(0^-) = 1 + R_\infty = \left(\Omega_0 + \sigma \Omega_1 + \sigma^2 \Omega_2
		+ \ldots\right)_{x=X=0}
		\label{eqn3.8}
	\end{equation}
	and
	\begin{equation}
		\Omega'(0^-) = ik_0(1 - R_\infty) = 
		\left( \frac{\partial}{\partial x} + \sigma^2 
		\frac{\partial}{\partial X} \right) \left( \Omega_0 + \sigma \Omega_1 + \sigma^2 \Omega_2
		+ \ldots \right)_{x=X=0}.
		\label{eqn3.9}
	\end{equation}
	At leading order, $\Omega_0$ satisfies the same wave equation (\ref{eqn3.2}) as in 
	$x < 0$ and its general solution is
	\begin{equation}
		\Omega_0(x,X) = A(X) \exp^{\ci k_0 x} + B(X) \exp^{-\ci k_0 x}.
		\label{eqn3.10}
	\end{equation}
	This implies that the leading order solution is not explicitly 
	dependent on individual realisations, $r(x)$; $A$ and $B$ will contain
	information relating to the statistical properties of $r(x)$ however.
	We require that long-scale variations, $A(X)$ and $B(X)$, tend to zero 
	as $X \to \infty$, whilst $A(0) = 1$ and $B(0) = R_\infty$ are determined from 
	the matching conditions (\ref{eqn3.8}), (\ref{eqn3.9}) at leading order.
	
	Since $|R_\infty| = 1$ there must be no net time-averaged transport of 
	energy flux in $x > 0$ and so we expect that
	\begin{equation}
		|A(X)| = |B(X)|.
		\label{eqn3.11}
	\end{equation}
	
	At $O(\sigma)$ we have
	\begin{equation}
		\frac{\partial^2 \Omega_1}{\partial x^2}  +  k_0^2 \Omega_1 = 
		-C_1
		\frac{\partial}{\partial x} \left( r(x)
		\frac{\partial \Omega_0}{\partial x} \right).
		\label{eqn3.12}
	\end{equation}
	Its solution can be determined using the Green's function for the
	one-dimensional wave equation,
	\begin{equation}
		g(x,x') = \frac{\exp^{\ci k_0 |x-x'|}}{2 \ci k_0},
		\label{eqn3.13}
	\end{equation}
	satisfying
	\begin{equation}
		\frac{\partial^2}{\partial x^2} g + k_0^2 g = \delta(x-x'),
		\label{eqn3.14}
	\end{equation}
	and outgoing as $|x-x'| \to \infty$. The right-hand side of (\ref{eqn3.12})
	is comprised of two terms forced by right- and left-propagating waves
	and the solution $\Omega_1$, in $x > 0$, is a superposition of solutions
	derived using $g$ and $\overline{g}$ (where the overbar denotes complex conjugate), respectively, in Green's identity 
	with the two components of $\Omega_1$ over $x > 0$ and results in
	\begin{multline}
		\Omega_1(x,X) = - \ci k_0 C_1 A(X) \int_0^\infty g(x,x') 
		\frac{\partial}{\partial x'} \left( r(x')
		\exp^{\ci k_0 x'} \right) \, \wrt x'
		\\
		+ \ci k_0 C_1 B(X) \int_0^\infty \overline{g}(x,x') 
		\frac{\partial}{\partial x'} \left( r(x')
		\exp^{-\ci k_0 x'} \right) \, \wrt x',
		\qquad x > 0.
		\label{eqn3.15}
	\end{multline}
	The use of $\overline{g}$ is non-standard and implies that the
	component of the first-order solution associated with left-propagating leading-order wave is represented by a distribution of incoming waves. This is required to satisfy the energy balance equation (\ref{eqn3.11}). Put another way, we require the amplitude, $B(X)$, of the left-going wave
	to grow as it propagates from right to left, its associated energy being 
	generated from the energy lost to outgoing waves from the 
	right-propagating wave with amplitude $A(X)$.
	
	Integrating by parts once, using $r(0) = 0$ (since the
	random variations in the bed or the ice continuously joins the 
	constant value set in $x < 0$) gives
	\begin{multline}
		\Omega_1(x,X) = - \ci k_0 C_1 A(X) \int_0^\infty
		\frac{\partial}{\partial x} g(x,x') r(x')
		\exp^{\ci k_0 x'} \wrt x'
		\\
		+  \ci k_0 C_1 B(X) \int_0^\infty
		\frac{\partial}{\partial x} \overline{g}(x,x') r(x')
		\exp^{-\ci k_0 x'} \wrt x'.
		\label{eqn3.16}
	\end{multline}
	Here $\partial_x g = - \partial_{x'} g$ has been used and we
	note that this function is discontinuous at $x=x'$.
	
	We also remark that $\Omega_1$ is a random function with zero mean since 
	$\langle \Omega_1 \rangle = 0$ follows from ensemble averaging (\ref{eqn3.16}) 
	and using (\ref{eqn2.3}).
	
	At $O(\sigma^2)$ we have
	\begin{equation}
		\frac{\partial^2 \Omega_2}{\partial x^2}  +  k_0^2 \Omega_2 = 
		-C_1 \frac{\partial}{\partial x} \left( r(x)
		\frac{\partial \Omega_1}{\partial x} \right)
		- 2 \frac{\partial^2 \Omega_0}{\partial x \partial X}
		+ \frac{\partial}{\partial x} \left( ( C_2r^2(x)
		+ C_3 r'^2(x)) 
		\frac{\partial \Omega_0}{\partial x} \right).
		\label{eqn3.18}
	\end{equation}
	We ensemble average the equation using the results from (\ref{eqn2.3}) and
	$\langle r'^2 \rangle = 2/\Lambda^2$ (this can be established 
	using the definition of the derivative as a limit) to give
	\begin{multline}
		\frac{\partial^2}{\partial x^2} \langle \Omega_2 \rangle +  k_0^2 
		\langle \Omega_2 \rangle = 
		- C_1
		\frac{\partial}{\partial x} \left\langle r(x)
		\frac{\partial \Omega_1}{\partial x} \right\rangle
		- 2 \ci k_0 (A'(X) \exp^{\ci k_0 x} - B'(X) \exp^{-\ci k_0 x} )
		\\
		-k_0^2 ( C_2 + 2 C_3/\Lambda^2)(A(X) \exp^{\ci k_0 x} + B(X)
		\exp^{-\ci k_0 x} ).
		\label{eqn3.19}
	\end{multline}
	It is instructive to write $\Omega_1$ from (\ref{eqn3.16}) in terms of separate wave-like components as
	\begin{multline}
		\Omega_1(x,X) = 
		-\frac{C_1 A(X) \ci k_0}{2} \left[ 
		\exp^{\ci k_0 x} \int_0^x r(x') \, \wrt x' - 
		\exp^{-\ci k_0 x} \int_x^\infty r(x') \exp^{2 \ci k_0 x'} \, \wrt x'
		\right]
		\\
		+ \frac{C_1 B(X) \ci k_0}{2} \left[ 
		\exp^{-\ci k_0 x} \int_0^x r(x') \, \wrt x' - 
		\exp^{\ci k_0 x} \int_x^\infty r(x') \exp^{-2 \ci k_0 x'} \, \wrt x'
		\right].
		\label{eqn3.17}
	\end{multline}
	We note that the leading-order right-propagating wave 
	excites both right-propagating waves which accumulate from interactions 
	with the bed to the left of the observation point, $x$, and left-propagating
	waves which represent the accumulation of upwave reflections from bed 
	interactions to the right of the observation point. Similar comments apply 
	to terms proportional to the leading-order left-propagating wave. 
	The ensemble averaging 
	of the first and third terms of (\ref{eqn3.17}) in (\ref{eqn3.19}) lead to a contribution to the attenuation which we describe as ``fictitious decay''. That is to say, it is a feature of wave scattering not experienced by individual waves, but which instead originates from phase cancellations from first-order waves when averaged over realisations of $r(x)$. The coefficient multiplying the two $\exp^{\pm \ci k_0 x}$ terms under scrutiny is a real integral which depends only on $r(x)$, the geometry, and hence randomness does not alter the phase of these contributions. This contrasts with the second and fourth terms in (\ref{eqn3.17}) which correspond to the accumulation of waves that have propagated from the field point $x$ to a point $x'$ and reflected by the bathymetry/broken ice $r(x')$ necessarily encoding randomness into the phase of these contributions. For the purpose of computing the attenuation experienced by individual waves we remove this fictitious decay effect, replacing (\ref{eqn3.17}) by
	\begin{equation}
		\Omega_1(x,X) = 
		\frac{C_1 A(X) \ci k_0}{2}  
		\exp^{-\ci k_0 x} \int_x^\infty r(x') \exp^{2 \ci k_0 x'} \, \wrt x'
		- \frac{C_1 B(X) \ci k_0}{2} 
		\exp^{\ci k_0 x} \int_x^\infty r(x') \exp^{-2 \ci k_0 x'} \, \wrt x'.
		\label{eqn3.17a}
	\end{equation}
	The only term requiring attention now is the first term on the
	right-hand side of (\ref{eqn3.19}) where $\Omega_1$ is given by (\ref{eqn3.17a}).
	It is straightforward to determine from (\ref{eqn3.17a}) that
	\begin{multline}
		\left\langle r(x) 
		\frac{\partial \Omega_1}{\partial x}
		\right\rangle 
		=
		-\frac{\ci k_0}{2} C_1 A(X) \exp^{\ci k_0 x} + k_0^2 C_1 A(X)
		\exp^{\ci k_0 x} \int_0^\infty
		\exp^{-\xi^2/\Lambda^2} \exp^{2 \ci k_0 \xi} \, \wrt \xi
		\\ 
		+ \frac{\ci k_0}{2} C_1 B(X) \exp^{-\ci k_0 x} + k_0^2 C_1 B(X)
		\exp^{-\ci k_0 x} \int_0^\infty
		\exp^{-\xi^2/\Lambda^2} \exp^{-2 \ci k_0 \xi} \, \wrt \xi
		\label{eqn3.21}
	\end{multline}
	after using the definition in (\ref{eqn2.4}) and making a substitution $\xi = x-x'$. 
	As demanded by (\ref{eqn3.19}), we need to take a further derivative which results in
	\begin{equation}
		C_1 \frac{\partial}{\partial x} \left\langle r(x) 
		\frac{\partial \Omega_1}{\partial x} \right\rangle = 
		\frac{C_1^2 k_0^2}{2} \big( A(X) F \exp^{\ci k_0 x} + 
		B(X)  \overline{F} \exp^{-\ci k_0 x} \big)
		\label{eqn3.23}
	\end{equation}
	where
	\begin{equation}
		F = 1 +\ci k_0 \int_{0}^\infty
		\exp^{-\xi^2/\Lambda^2} \exp^{2 \ci k_0 \xi} \, \wrt \xi =
		1 + \frac{\sqrt{\pi}}{2} \ci k_0 \Lambda \,
		\exp^{-k_0^2 \Lambda^2}
		(1 + \ci \, \mbox{erfi}(k_0 \Lambda)),
		\label{eqn3.24}
	\end{equation}
	(see, e.g., \citet{Mei_Li_2004}) and $\mbox{erfi}(\cdot)$ is the imaginary error function.
	
	Armed with (\ref{eqn3.23}), we return to the governing equation (\ref{eqn3.19}) for $\langle \Omega_2 \rangle$ and note that
	the right-hand side contains secular
	terms; that is functions proportional to $\exp^{\pm \ci k_0 x}$.
	These must be removed to avoid unbounded growth in the solution
	for $\langle \Omega_2 \rangle$ as $x \to \infty$. In 
	other words we wish to obtain
	\begin{equation}
		\frac{\partial^2}{\partial x^2} \langle \Omega_2 \rangle +  k_0^2 
		\langle \Omega_2 \rangle = 0,
		\label{eqn3.29}
	\end{equation}
	requiring $A(X)$ and $B(X)$ to satisfy the solvability conditions
	\begin{equation}
		2 \ci k_0 A'(X) = -k_0^2 A(X) \left( C_1^2 \left( \frac{1}{2} + 
		\frac{\sqrt{\pi}}{4} \ci k_0 \Lambda 
		\exp^{-k_0^2\Lambda^2}(1 + \ci \, \mbox{erfi}(k_0 \Lambda) \right)
		+ C_2 + 2 C_3/\Lambda^2 \right)
		\label{eqn3.30}
	\end{equation}
	and
	\begin{equation}
		-2 \ci k_0 B'(X) = -k_0^2 B(X) \left( C_1^2 \left( \frac{1}{2} -
		\frac{\sqrt{\pi}}{4} \ci k_0 \Lambda \exp^{-k_0^2\Lambda^2}(1 - \ci \, \mbox{erfi}(k_0 \Lambda) \right)
		+ C_2 + 2 C_3/\Lambda^2 \right).
		\label{eqn3.31}
	\end{equation}
	Solving for $A(X)$ with $A(0) = 1$ gives
	\begin{equation}
		A(X) = \exp^{ - Q X + \ci \kappa X}
		\label{eqn3.32}
	\end{equation}
	where
	\begin{equation}
		Q = \frac{\sqrt{\pi}}{8} C_1^2 k_0^2 \Lambda \exp^{-k_0^2 \Lambda^2}
		\label{eqn3.33}
	\end{equation}
	and
	\begin{equation}
		\kappa  = C_1^2 \left( \frac{k_0}{4} - \frac{\sqrt{\pi}}{8} k_0^2 \Lambda
		\exp^{-k_0^2 \Lambda^2} \mbox{erfi}(k_0 \Lambda)\right) + k_0 C_2/2 + k_0 C_3/\Lambda^2.
		\label{eqn3.34}
	\end{equation}
	Meanwhile, solving (\ref{eqn3.31}) for $B(X)$ with $B(0) = R_\infty$ 
	such that $|R_\infty| = 1$ gives
	\begin{equation}
		B(X) = R_\infty \exp^{ - Q X - \ci \kappa X}
		\label{eqn3.35}
	\end{equation}
	and thus (\ref{eqn3.16}) is satisfied.
	
	Had the first and third terms in (\ref{eqn3.17}) not
	been removed and (\ref{eqn3.17}) not been replaced by
	(\ref{eqn3.17a}) then, amongst other changes, the expression in (\ref{eqn3.33}) would have
	have been replaced by $Q = (\sqrt{\pi}/{8}) C_1^2 k_0^2 \Lambda (1+\exp^{-k_0^2 \Lambda^2})$. A similar attenuation factor
	is determined in the work of 
	\citet[Section~7.4]{Mei_Stiassnie_P._2005} and \citet{Bennetts_Peter_Chung_2014}. The additional factor of $+1$, associated with phase cancellation in the ensemble averaging, completely changes the character of attenuation.
	\citet{Bennetts_Peter_Chung_2014} highlight the discrepancy
	between theoretical results and attenuation measured through
	discrete numerical simulations, most notably in Figures 5 and 6 of their paper.
	Moreover, the expression for $B(X)$ 
	would also change with the factor of $Q$ associated with (\ref{eqn3.35}) replaced by $Q = (\sqrt{\pi}/{8}) C_1^2 k_0^2 \Lambda (-1+\exp^{-k_0^2 \Lambda^2})$ implying exponential growth towards infinity of the left-propagating wave whilst (\ref{eqn3.16}) is no longer satisfied.
	
	Returning to (\ref{eqn3.10}) gives
	the leading order solution in $x > 0$ as
	\begin{equation}
		\Omega(x) \approx \Omega_0(x,\sigma^2 x) = \exp^{- \sigma^2 Q x} 
		\left( \exp^{\ci (k_0 + \sigma^2 \kappa) x}
		+ R_\infty \exp^{-i (k_0 + \sigma^2 \kappa) x} \right).
		\label{eqn3.40}
	\end{equation}
	Furthermore, since $\langle \Omega_1 \rangle = 0$, corrections to (\ref{eqn3.40}) 
	are $O(\sigma^2)$. From (\ref{eqn3.40}) the attenuation 
	rate is defined to be
	\begin{equation}
		k_i = \sigma^2 Q = 
		\frac{\sqrt{\pi}}{8} k_0^2 \sigma^2 \Lambda C_1^2 \exp^{-k_0^2 \Lambda^2}
		\label{eqn3.41}
	\end{equation}
	with $C_1$ given by (\ref{eqn3.4}) (or (\ref{eqn3.4b})), a factor which depends upon $k_0h_0$ (and $d_0/h_0$). In the case of a randomly-varying bed with no ice cover and assuming $C_1^2 \approx 1$ since $Kh_0 \ll 1$, the maximum value of $k_i$ will occur at $k_0 \Lambda \approx 1$. This value can be interpreted as being associated with
	Bragg resonance which occurs close to $k_0 \Lambda = 1$ for periodic beds with periodicity $\Lambda$. Bragg resonance is characterised by coherent multiple reflections. 
	In the case of varying ice $C_1^2\approx d_0^2/(h_0-d_0)^2$ which alters the 
	magnitude of the attenuation, but not the condition $k_0\Lambda \approx 1$ for the
	maximum.
	
	For $k_0 \Lambda \ll 1$, $k_i \propto 
	k_0^2$ and whilst for $k_0 \Lambda \gg 1$ the attenuation decays 
	exponentially as $k_0 \Lambda$ increases, although we note this limit is outside the long wavelength assumptions used to develop this model. The latter result holds
	in this long wavelength model and contrasts with the conclusions drawn by previous researchers (e.g. see 
	\citet{Devillard_Dunlop_Souillard_1988}, 
	\citet[Section~5]{Mei_Stiassnie_P._2005}) who associate
	exponential decay in wave attenuation as a finite water depth effect.
	
	These conclusions are based on a long wave model of wave propagation with randomness described by a continuously varying function. For short wave scattering by floating broken ice, for example, the physics will be different as scattering by discrete ice floes will need to be correctly modelled.
	
	\section{Numerical methods and simulations}\label{Section 5}
	
	\subsection{Generating a random surface}\label{Section 5a}
	
	In order to numerically generate a random function, $r(x)$, with statistical properties (\ref{eqn2.3}) and (\ref{eqn2.4}) characterised by the RMS height $1$ and the correlation length $\Lambda$ we implement the weighted moving average method described in \citet{sarris_2021_attenuation} and originally due to \citet{ogilvy_1988_computer}.
	The function $r(x)$ will be defined at $x = x_i = i \Delta x$ for $i=0,\ldots, V$ where $\Delta x = L/V$; either $\Delta x$ or $V$ can be used as the numerical parameter defining the resolution of the random surface.
	
	We generate the Gaussian weights
	\begin{equation}
		w_j = W \exp^{-2(j\Delta x)^2/\Lambda^2}
		\label{eqn4.1}
	\end{equation}
	for $j=-M,\ldots,M$ where $M = \lfloor 4\Lambda/(\Delta x\sqrt{2}) \rceil$
	(denoting integer part) is a truncation parameter and 
	$W$ is defined to normalise these values so that
	\begin{equation}
		\sum_{j=-M}^{M} w_j = 1.
		\label{eqn4.2}
	\end{equation}
	Next, we define
	\begin{equation}
		\sigma_v^2 = 1/\sum_{j=-M}^{M}w_j^2
		\label{eqn4.3}
	\end{equation}
	which is used to generate the $2N+1$ uncorrelated random numbers $v_i$,
	$-N \leq i \leq N$ from a Gaussian distribution with a variance of $\sigma_v$.
	The height of a random surface at $x=x_i$ is defined by
	\begin{equation}
		r_i = \sum_{j=-M}^{M} w_j v_{j+i+M-N}, 
		\qquad i=0,\ldots,V
		\label{eqn4.4}
	\end{equation}
	requiring $N$ to be defined by $2N = V+2M$. Our theory requires that
	$r(x) = 0$ at $x=0$, $x=L$ and that these values are approached smoothly from within the interval $x \in (0,L)$. We thus introduce a Tukey 
	smoothing window at either end of the interval of length $\Lambda$ (assumed to be
	less than $L/2$) via
	\begin{equation}
		r(x_i) = \left\{ \begin{array}{l}
			{\displaystyle r_i, \qquad V_\Lambda +1 \leq i \leq V-V_\Lambda -1},
			\\ \noalign{\vskip4pt}
			{\displaystyle r_i \left( \frac12 - \frac12 \cos \left( \frac{i\pi}{V_\Lambda} \right) \right), \qquad i=0,\ldots,V_{\Lambda}},
			\\ \noalign{\vskip4pt}
			{\displaystyle r_i \left( \frac12 - \frac12 \cos \left( \pi \frac{V-i}{V_\Lambda} \right) \right), \qquad i=V-V_{\Lambda},\ldots,V},
		\end{array} \right.
		\label{eqn4.5}
	\end{equation}
	where $V_\Lambda = \lfloor \Lambda/\Delta x \rceil$. Numerically, we ensure $V_\Lambda$, which represents the number of points per characteristic length of bed, is sufficiently large.
	
	\subsection{Determining decay via a transfer matrix}\label{Section 5b}
	
	Simulations of scattering are performed over a region $0 < x < L$ with $L/h_0 \gg 1$. Taking $L$ to be large is done since we wish to compare are results with the theoretical results where $L= \infty$. Thus, we aim to ensure that waves pass over enough of the bed for the effect of randomness to be felt. Attenuation over longer beds can also help suppress multiple scattering effects associated with the junctions at $x=0$ and $x=L$ between constant and random surfaces. However, the method described below for determining attenuation is insensitive to multiple scattering effects.
	
	Instead of (\ref{eqn1.7}), (\ref{eqn1.8}), let us momentarily express the solution in $x < 0$, $x > L$ more generally as
	\begin{align}
		\Omega(x)=\begin{cases}
			A_{-}\exp^{\ci k_0x} + B_{-}\exp^{-\ci k_0x},	&x<0\\
			A_{+}\exp^{\ci k_0x} + B_{+}\exp^{-\ci k_0x},	&x>L
		\end{cases}
		\label{eqn4.6}
	\end{align}
	for complex constants $A_\pm$, $B_\pm$, representing amplitudes of right- and left-propagating waves, respectively, whilst $k_0$ satisfies (\ref{eqn1.9}).
	
	We encode scattering using either a $2 \times 2$ scattering matrix, ${\sf S}$, 
	satisfying
	\begin{equation}
		\begin{pmatrix}
			A_+ \\
			B_- 
		\end{pmatrix}={\sf S} \begin{pmatrix}
			A_- \\
			B_+ 
		\end{pmatrix}
		\label{eqn4.7}
	\end{equation}
	which relates outgoing to incoming waves or a $2 \times 2$ transfer matrix, ${\sf P}$, satisfying
	\begin{equation}
		\begin{pmatrix}
			A_+ \\
			B_+ 
		\end{pmatrix}= {\sf P} \begin{pmatrix}
			A_- \\
			B_- 
		\end{pmatrix}
		\label{eqn4.8}
	\end{equation}
	which relates waves in $x > L$ to waves in $x < 0$.
	Energy conservation requires incoming and outgoing wave energy fluxes balance so that $|A_-|^2 + |B_+|^2 = |A_+|^2 + |B_-|^2$ and this implies $\overline{\sf S}^T {\sf S} = {\sf I}$ where ${\sf I}$ is the Identity and the overbar denotes conjugation; ${\sf S}$ is a unitary matrix. Multiplying (\ref{eqn4.8}) by $(\overline{A}_+, - \overline{B}_+)^T$  results in a similar identity 
	\begin{equation}
		{\sf E} \overline{\sf P}^T {\sf E} {\sf P} = {\sf I}, \qquad 
		{\sf E} = \begin{pmatrix} 1 & 0 \\ 0 & -1 \end{pmatrix}.
		\label{eqn4.9}
	\end{equation}
	This is sufficient to show that if $\lambda$ is an eigenvalue of ${\sf P}$ then so is $\overline{\lambda}$, as is $1/\overline{\lambda}$. The pair of eigenvalues $\lambda_\pm$ of ${\sf P}$ are therefore either both real, occurring in reciprocal pairs, or complex conjugates lying on the unit circle.
	
	As shown in, for example \citet{porandpor03}, the eigenvalues characterise wave propagation across $0 < x < L$: if $\lambda_{\pm}$ are complex conjugates then there is no attenuation as waves travel from left to right. If, however, $\lambda_{\pm}$ are real, then writing $\lambda_+=\exp^{-k_i  L}$ and $\lambda_-=\exp^{k_i  L}$, say, indicate that right- and left-propagating waves are attenuated with the rate $k_i $.
	
	Since the transfer matrix, $P$, describes the solution over $0 < x < L$ without coupling to the solution in $x < 0$ and $x > L$ its eigenvalues determine decay (or otherwise) without interference from multiple scattering effects associated with waves being reflected at the junctions $x=0$ and $x=L$.
	
	The entries of ${\sf S}$ and ${\sf P}$ requires us to solve (\ref{eqn1.4}). We follow \citet{porter_2019}, 
	write $x = \xi L$, $p(\xi) = \Omega(x) = (1+R)p_1(\xi) + \im k_0\doublehat{d_0}(1-R)p_2(\xi)$ and $q(\xi) = \doublehat{d}(x)\Omega'(x)= (1+R)q_1(\xi) + \im k_0\doublehat{d_0}(1-R)q_2(\xi)$, where $\doublehat{d_0} = \doublehat{d}(0)$, and numerically solve the dimensionless coupled first order system
	\begin{equation}
		p_i'(\xi) = (L/\doublehat{d}(L \xi)) q_i(\xi), \qquad q'_i(\xi) = -KL p_i(\xi), \qquad 0 < \xi < 1
		\label{eqn4.10}
	\end{equation}
	for $i=1,2$ with the initial conditions $p_1(0) = 1$, $q_1(0) = 0$ and $p_2(0) = 0$ and $q_2(0) = 1$. This allows us, after matching to the solution given by (\ref{eqn4.6}) in $x < 0$ and $x > L$ and with some manipulation of the algebra, to express the solution either using (\ref{eqn4.7}) with
	\begin{equation}
		{\sf S} = \begin{pmatrix}
			\ci \doublehat{d_0}k_0 p_2(1) - p_1(1) & \exp^{\ci k_0 L} \\
			\ci \doublehat{d_0}k_0 q_2(1) - q_1(1) & \ci \doublehat{d_0}k_0 \exp^{\ci k_0 L}
		\end{pmatrix}^{-1}
		\begin{pmatrix}
			\ci \doublehat{d_0}k_0 p_2(1) + p_1(1) & \exp^{-\ci k_0 L} \\
			\ci \doublehat{d_0}k_0 q_2(1) + q_1(1) & -\ci \doublehat{d_0}k_0 \exp^{-\ci k_0 L}
		\end{pmatrix}
		\label{eqn4.11}
	\end{equation}
	or using (\ref{eqn4.8}) with
	\begin{equation}
		{\sf P} =	\begin{pmatrix}
			\exp^{\ci k_0L} & \exp^{-\ci k_0L} \\
			\ci \doublehat{d_0}k_0 \exp^{\ci k_0L} & -\ci \doublehat{d_0}k_0 \exp^{-\ci k_0L}
		\end{pmatrix}^{-1}
		\begin{pmatrix}
			\ci\doublehat{d_0}k_0p_2(1)+p_1(1) & -\ci \doublehat{d_0}k_0p_2(1)+p_1(1) \\
			\ci \doublehat{d_0}k_0q_2(1)+q_1(1) & -\ci \doublehat{d_0}k_0q_2(1)+q_1(1) 
		\end{pmatrix}.
		\label{eqn4.12}
	\end{equation}
	When we set $A_- = 1$ and $B_+=0$, $B_- = R_L$ and $A_+ = T_L$ become the reflection
	and transmission coefficients to due waves incident from $x < 0$ which are most easily determined from (\ref{eqn4.7}) with (\ref{eqn4.11}).
	
	Attenuation, on the other hand, simply requires us to evaluate the pair of eigenvalues of ${\sf P}$ from (\ref{eqn4.12}). The corresponding decay rate is then determined from $k_i  = | \ln |\lambda_+||/L$ which, in the case of complex conjugate eigenvalues is zero.
	
	For the ensemble averaging the results we run $N \gg 1$ simulations of different realisations of the bed or the ice thickness and then compute
	\begin{equation}
		\langle k_i  \rangle = \frac1N \sum_{n=1}^{N} k_i ,
		\qquad
		\langle |R_L| \rangle = \frac1N \sum_{n=1}^{N} |R_L|,
		\qquad
		\langle |T_L| \rangle = \frac1N \sum_{n=1}^{N} |T_L|,
		\label{eqn4.13}
	\end{equation}
	where the terms under the sum represent the output of each random simulation.
	Depending on numerical parameters used, computations of the three averages will typically take between 20 and 200 seconds on a standard desktop PC when $N=500$.
	A standard Runge-Kutta-Fehlberg method is used to solve (\ref{eqn4.10}).
	
	\section{Results for randomly varying beds without ice cover}\label{Section 6}
	
\begin{figure}
		\centering
		\includegraphics[width=\linewidth,valign=t]{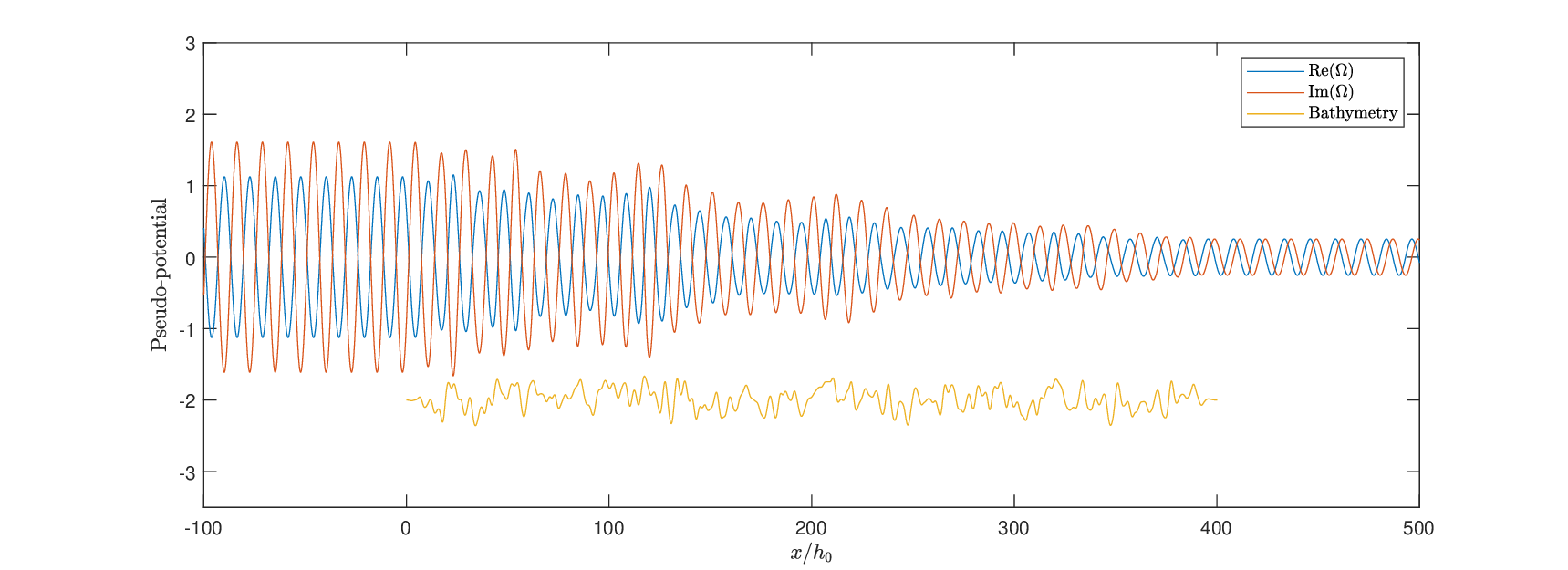}
		\caption{An example of the pseudo-potential (real and imaginary parts of $\Omega(x)$) and an overlay of the random function representing bathymetry $0 < x < L$. Here, $\sigma^2 = 0.02$, $\Lambda=2h_0$ and $L=400h_0$.}
		\label{fig2}
	\end{figure}

	Initially, we wish to comment that the following results only account for multiple-scattering effects present in shallow water and does not account for other non-negligible sources of attenuation such as bed friction and other sources of physical dissipation. We start by illustrating the numerical solution from a single realisation of a random bed. In Fig.~\ref{fig2} the function $h(x)/h_0$ is plotted
	about -2 on the vertical scale in the figure which is used to represent the real and imaginary parts of the pseudo-potential. In this simulation the bed is defined by $h_0 = 1$, $\Lambda = 2h_0$, $\sigma^2 = 0.02$ and $L = 400h_0$. The figure illustrates the randomness of the wave response over the bed and partial reflection and transmission of the incident wave. Note that partial transmission is not necessarily a result
	of wave attenuation over the random bed and occurs whenever
	there are changes in propagation characteristics. See, for example,
	the results of \citet{mei_black_1969} for wave propagation over a rectangular step. 
	
	We should also mention that the function describing the random beds are stored
	numerically at discrete points at a sufficiently high resolution that 
	linear interpolation can be used to accurately represent $h(x)$ and
	$h'(x)$ at any intermediate points needed by the numerical integration
	routine.
	
	\begin{figure}
		\centering
		\begin{subfigure}[t]{0.03\textwidth}
			\text{(a)}
		\end{subfigure}
		\begin{subfigure}[t]{0.45\textwidth}        \centering
			\includegraphics[width=\linewidth, valign=t]{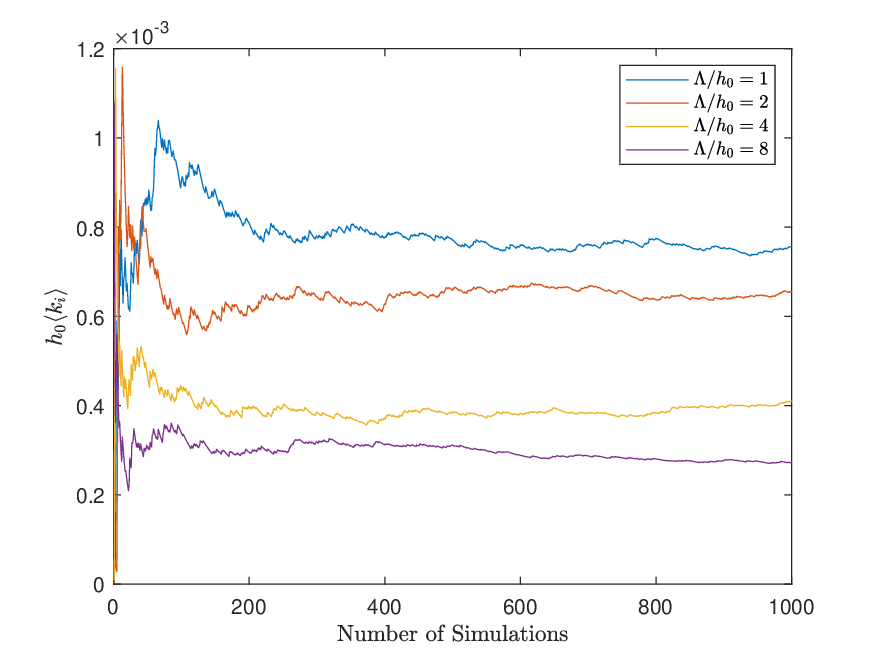}
		\end{subfigure}\hfill
		\begin{subfigure}[t]{0.03\textwidth}
			\text{(b)}
		\end{subfigure}
		\begin{subfigure}[t]{0.45\textwidth}        \centering
			\includegraphics[width=\linewidth, valign=t]{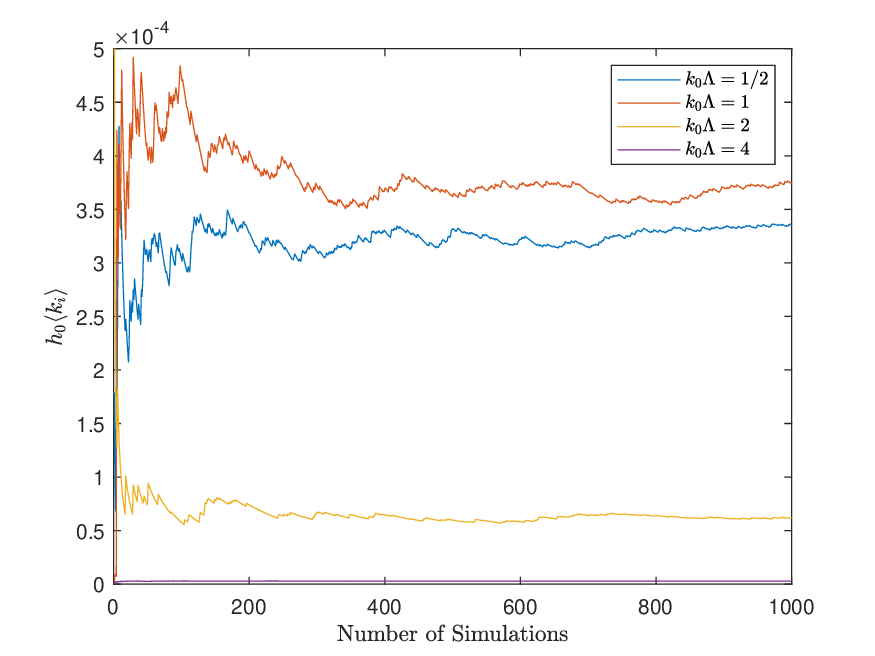}
		\end{subfigure}
		\caption{The variation of the dimensionless attenuation constant as $N$, the number of simulations, increases for random bathymetry with $L=400h_0$ and $\sigma^2 = 0.02$. In (a) $k_0\Lambda=1$ is fixed and $\Lambda/h_0$ is varied; in (b) $\Lambda/h_0 = 4$ is fixed and $k_0 \Lambda$ is varied.}
		\label{fig3}
	\end{figure}
	
	In Figure \ref{fig3} we present plots illustrating the typical
	convergence of the dimensionless attenuation rate, $h_0 \langle k_i  \rangle$, against $N$, the number of simulations. In both plots, the bed is of fixed length of $L = 400h_0$ with vertical variations parametrised by 
	$\sigma^2 = 0.02$. In one plot we fix frequency at $k_0 \Lambda = 1$ 
	and vary $\Lambda/h_0 = 1,2,4,8$. In the second plot we fix 
	$\Lambda/h_0 = 4$ and vary $k_0 \Lambda = 0.5,1,2,4$. Similar results
	are found when $\sigma$ is varied with $\Lambda/h_0$ and $k_0 \Lambda$
	are held fixed. These and other tests performed suggest $N=500$ simulations is sufficiently large to obtain reasonable convergence to the ensemble average when balanced against computational time. We use $N=500$ by default occasionally increasing $N$ when
	there is good reason to do so.
	Generally we find convergence is faster for larger $k_0 \Lambda$ and for larger $\Lambda/h_0$ and smaller values of $\sigma$.
	
	\begin{figure}
		\centering
		\includegraphics[width=0.5\linewidth,valign=t]{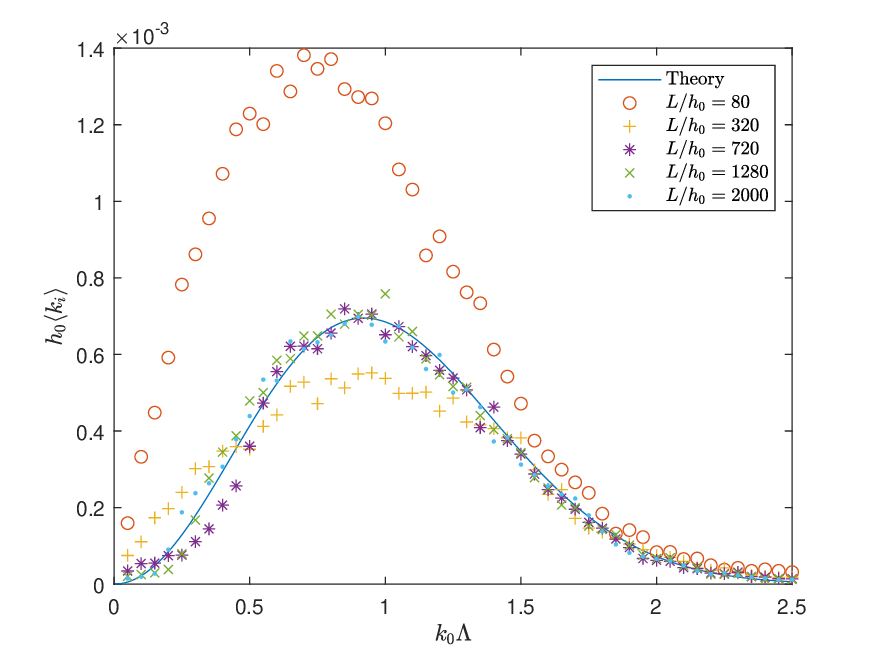}
		\caption{Non-dimensional ensemble averaged attenuation coefficient for $N=500$ simulations for beds of increasing length $L$, compared to theory. Here, $\sigma^2 = 0.02$ and $\Lambda=2h_0$.}
		\label{fig4}
	\end{figure}
	
	The next issue we address is the effect of bed length on convergence of the attenuation rate computed from the numerical simulation. In Fig.~\ref{fig4} we have fixed the bed statistics to $\sigma^2 = 0.02$, $\Lambda/h_0 = 2$ and plotted the ensemble average of dimensionless attenuation coefficient against $k_0 \Lambda$ for bed lengths increasing from $L = 80h_0$ to $2000h_0$. Overlaid is the theoretical prediction for a semi-infinite bed given by (\ref{eqn3.41}). Thus, in Fig.~\ref{fig4}, the numerical simulations
	appear to be converging to the theory as $L \to \infty$.
	
	\begin{figure}
		\centering
		\begin{subfigure}[t]{0.03\textwidth}
			\text{(a)}
		\end{subfigure}
		\begin{subfigure}[t]{0.45\textwidth}        \centering
			\includegraphics[width=\linewidth,valign=t]{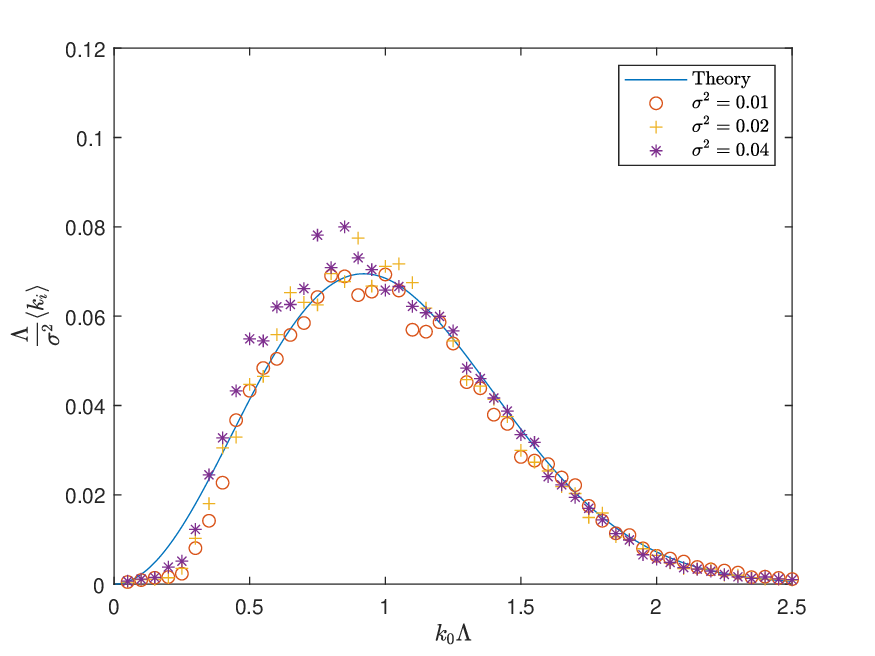}
		\end{subfigure}\hfill
		\begin{subfigure}[t]{0.03\textwidth}
			\text{(b)}
		\end{subfigure}
		\begin{subfigure}[t]{0.45\textwidth}        \centering
			\includegraphics[width=\linewidth,valign=t]{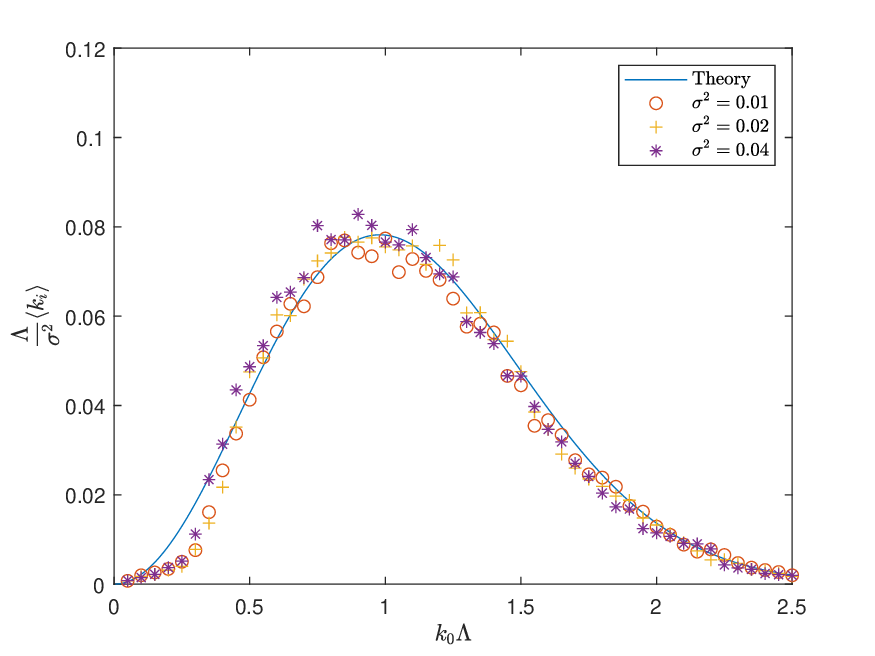}
		\end{subfigure}
		\caption{Scaled ensemble averaged attenuation coefficients for $N=500$ simulations for beds of length $L = 10\Lambda/\sigma^2$, compared with theory: (a) $\Lambda/h_0 = 2$, (b) $\Lambda/h_0 = 4$.}
		\label{fig5}
	\end{figure}
	
	Fig.~\ref{fig4} indicates that the section of variable bed needs to be sufficiently long for multiple wave scattering interactions over the variable bed to accurately capture decay due to randomness. Since this is determined by calculating $\lambda_{\pm} = \exp^{\mp k_i L}$ for each realisation, it is expected that $L$ will be defined by $k_i L = C$ for a constant $C$ sufficiently large that variations due to randomness in eigenvalues $\lambda_{\pm}$ of the transfer matrix ${\sf P}$ remain on the real line. Extensive numerical experimentation has indicated that the rule $k_i L = 1$, $k_i$ being the theoretically-derived attenuation rate, seem to produce ensemble averages which converge across all frequencies although a small proportion of realisations still return eigenvalues from the transfer matrix indicating no attenuation. However, setting $L$ according to the rule $k_i L = 1$ implies increasingly long beds in both the low- and high-frequency limits. Numerical simulations become both computationally expensive and prone to rounding errors. Instead we have produced results with $L = 10 \Lambda/\sigma^2$ which has the benefit of being independent of frequency so that the same bed realisations can be used across all frequencies. In doing so we are not able guarantee convergence of numerical results for $k_0 \Lambda$ such that $k_0 \Lambda \exp^{-k_0^2 \Lambda^2/2} \lesssim 0.05 \sqrt{\Lambda/\sigma^2 h_0}$. For example, with $\sigma^2 = 0.01$ and $\Lambda/h_0 = 2$ this translates to $k_0 \Lambda \lesssim 0.7$. Discrepancies between the numerical simulations and theory are noticeable at low frequencies especially for $\sigma^2 = 0.01$ in the plots in Fig.~\ref{fig5}. The issue of $L$ not being sufficiently large for high frequencies does not appear to affect the results so much. Similar general comments apply later to Fig.~\ref{fig10}, although we do notice the lack of convergence at high frequencies in the case where $L$ takes its lowest value.
	
	In Fig.~\ref{fig5} we collapse simulated data for different values of $\sigma^2 = 0.01,0.02,0.04$ onto the theoretical predictions for the scaled attenuation $\Lambda \langle k_i \rangle/\sigma^2$ for two values of $\Lambda/h_0 = 2,4$. The only differences in the two theoretical predictions are due to the scaling $C_1^2$ which
	depends on both $k_0 \Lambda$ and $\Lambda/h_0$. Although there is noise in the data, we have confirmed through extensive runs of the model that the fit between the data and the theory improves as $\sigma^2$ tends to zero. This is expected since the theoretical attenuation is a leading order result from an asymptotic expansion in $\sigma^2$.
	The numerical 
	results in Fig.~\ref{fig5} appear similar in character to results produced by \citet{Bennetts_Peter_Chung_2014} in their Figure 5 where they highlighted the discrepancy between decay 
	experienced by individual realisations and the decay predicted by their theory. These authors correctly
	surmise: ``{\it We deduce that the dominant source of attenuation of the effective wave elevation is wave cancellation (decoherence).}'' In our analysis, we identified and removed the terms which give rise to this ``fictitious decay''.
	
	\begin{figure}
		\centering
		\begin{subfigure}[t]{0.03\textwidth}
			\text{(a)}
		\end{subfigure}
		\begin{subfigure}[t]{0.45\textwidth}        \centering
			\includegraphics[width=\linewidth,valign=t]{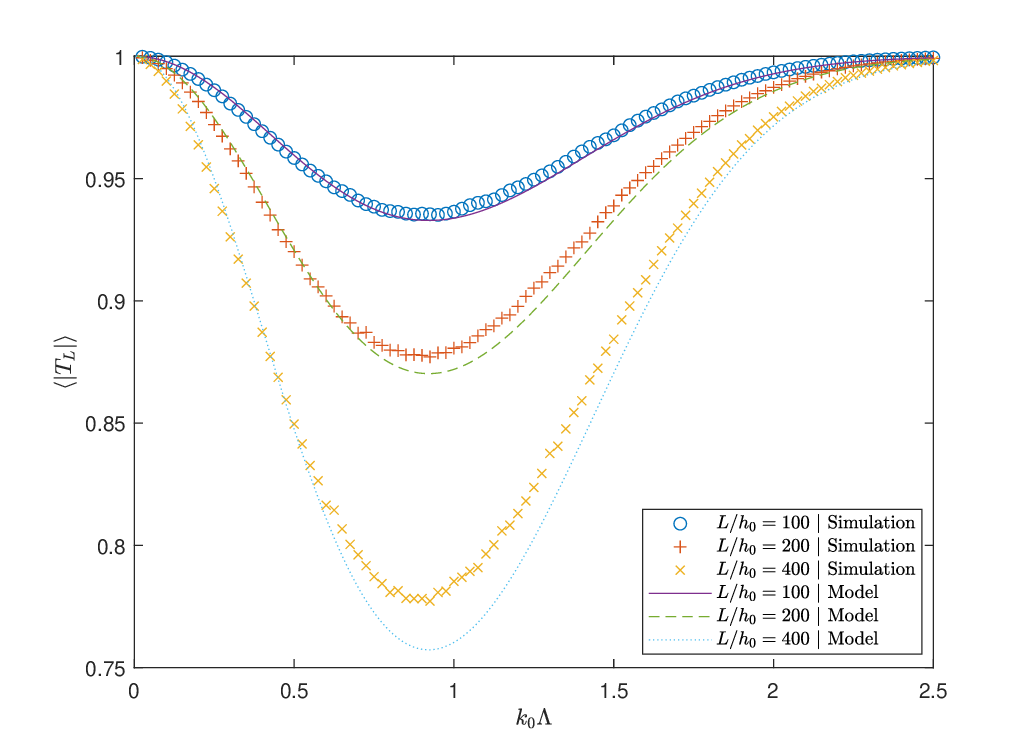}
		\end{subfigure}
		\begin{subfigure}[t]{0.03\textwidth}
			\text{(b)}
		\end{subfigure}
		\begin{subfigure}[t]{0.45\textwidth}        \centering
			\includegraphics[width=\linewidth,valign=t]{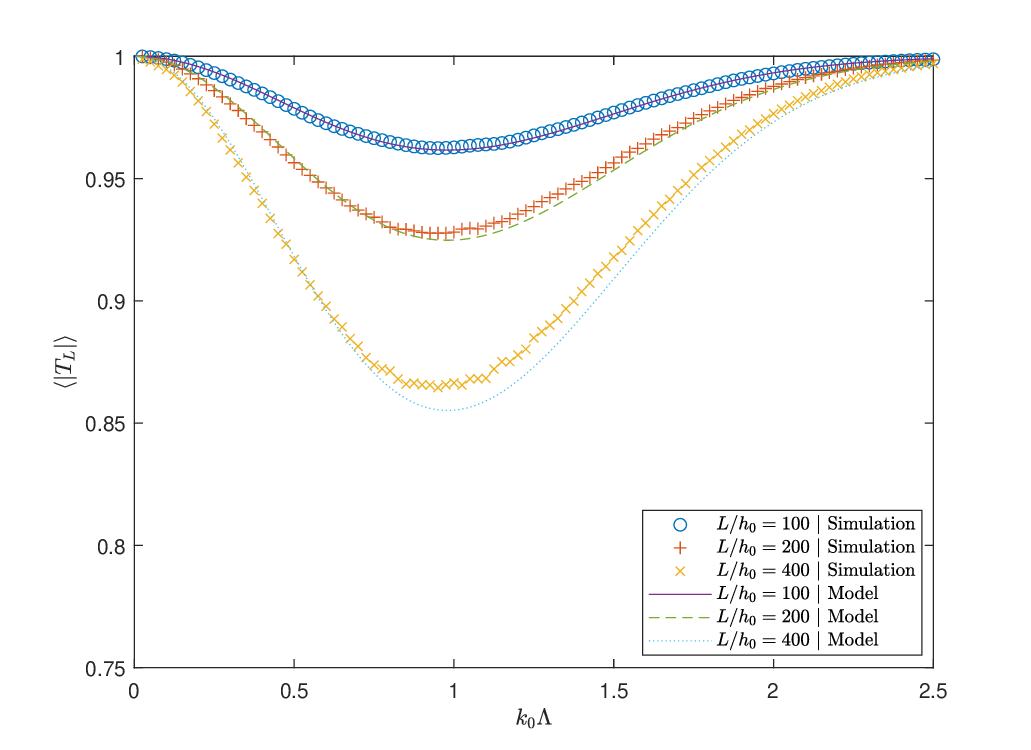}
		\end{subfigure}
		\caption{Variation with frequency of the ensemble average of the modulus of the transmission coefficient for $N=20000$ random bed simulations with statistical properties: (a) $\sigma^2 = 0.02, \Lambda = 2h_0$, (b) $\sigma^2 = 0.02, \Lambda = 4h_0$. Model refers to the curve fit $\langle |T_L| \rangle = \exp^{-{k_iL}}$.}
		\label{fig6}
	\end{figure}
	
	In Fig.~\ref{fig6} we show ensemble average of the modulus of the transmission coefficient against frequency for beds with statistics $\sigma^2 = 0.02$, $\Lambda/h_0 = 2$
	in one plot and $\Lambda/h_0 = 4$ in the second,
	for different lengths $L/h_0 = 100,200,400$. The limit
	$L \to \infty$ results in $T_\infty = 0$, so the convergence
	to this limit
	with increasing $L$ is slow and the variations with $L$ are
	significant. Results have been produced by averaging over
	20000 simulations to produce much more accurate averages
	than in previous results. This is done to give a clear
	indication of the fit between the numerical results for
	$\langle |T_L| \rangle$ for beds of finite length $L$
	and an approximate fit given by the curve $\langle |T_L| \rangle = 
	\exp^{-k_i L}$ where $k_i$ is the attenuation rate defined
	by (\ref{eqn3.41}) for a semi-infinite bed. 
	We offer no formal theoretical basis for this `model' fit, but note it agree with exact results in both limits $L\to 0$ and $L \to \infty$. Heuristically, this fit might be explained by the reflection at the junctions at $x=0$ and $x=L$ between varying and constant depths being weak in comparison to the accumulated attenuation via multiple-scattering over the length of random bed.
	
\begin{figure}
		\centering
		\begin{subfigure}[t]{0.03\textwidth}
			\text{(a)}
		\end{subfigure}
		\begin{subfigure}[t]{0.45\textwidth}        \centering
			\includegraphics[width=\linewidth,valign=t]{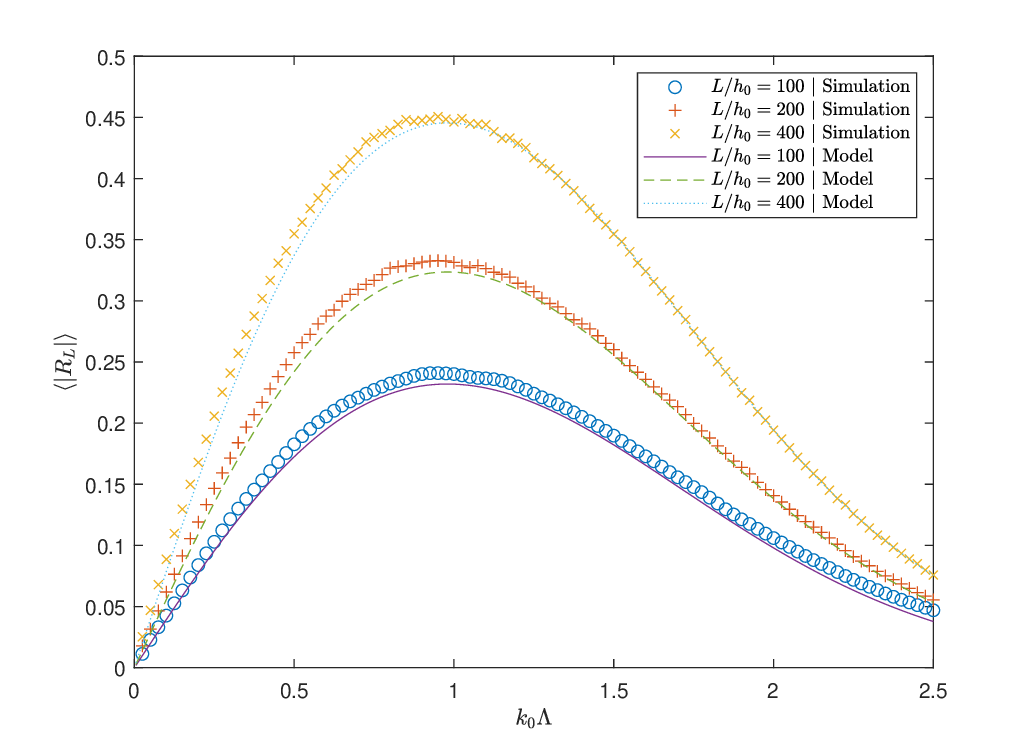}
		\end{subfigure}\hfill
		\begin{subfigure}[t]{0.03\textwidth}
			\text{(b)}
		\end{subfigure}
		\begin{subfigure}[t]{0.45\textwidth}        \centering
			\includegraphics[width=\linewidth,valign=t]{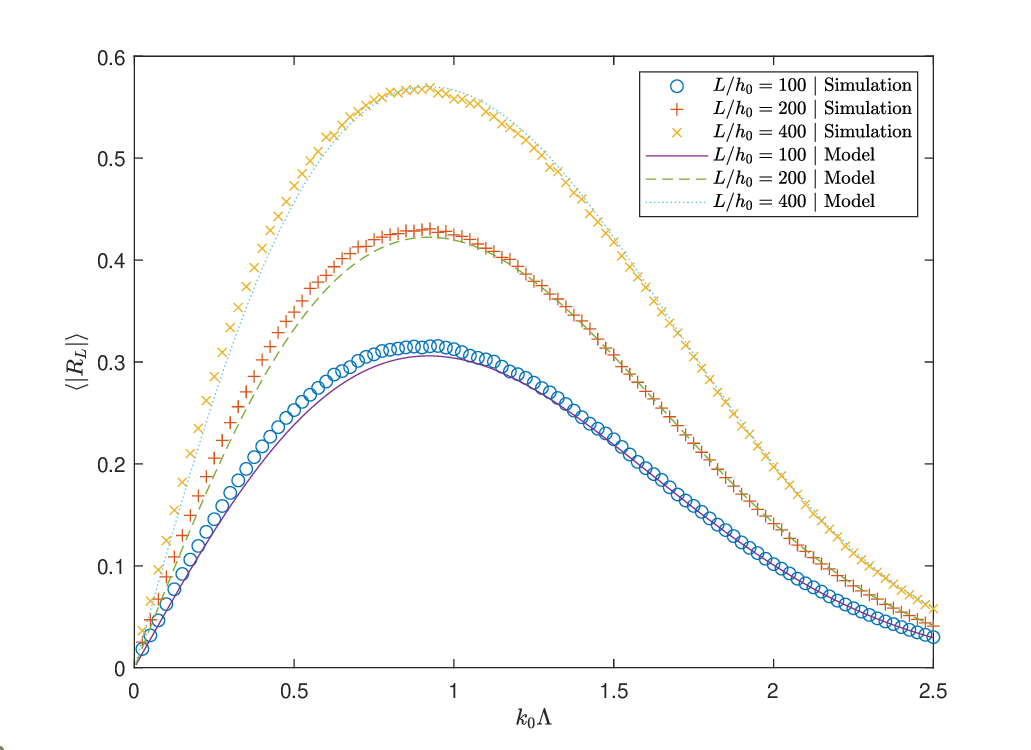}
		\end{subfigure}
		\caption{The ensemble average of the reflection coefficient for $N=20000$ simulations of random beds of varying length with statistics: (a) $\sigma^2 = 0.02, \Lambda = 2h_0$, (b) $\sigma^2 = 0.02, \Lambda = 4h_0$. The model fit are curves given by $\langle |R_L| \rangle = \sqrt{1 - \exp^{-{\sqrt{2}k_iL}}}$.}
		\label{fig7}
	\end{figure}

	Another model fit has been found for the ensemble average of the reflection coefficient for scattering over random beds of finite extent. These results are shown
	in Fig.~\ref{fig7} for beds of different
	lengths with $N=20000$ simulations used for averaging.
	The model fit $\langle |R_L| \rangle = \sqrt{1 - \exp^{-{\sqrt{2}k_iL}}}$ to these results has no theoretical
	basis but appears to be remarkably accurate. We felt it useful to present this result in the event that it might have practical use or help develop new theoretical results for scattering over random beds of finite extent.
	
	\section{Results for randomly varying ice thickness in water of constant depth}\label{Section 7}
	
	\begin{figure}
		\centering
		\includegraphics[width=\linewidth,valign=t]{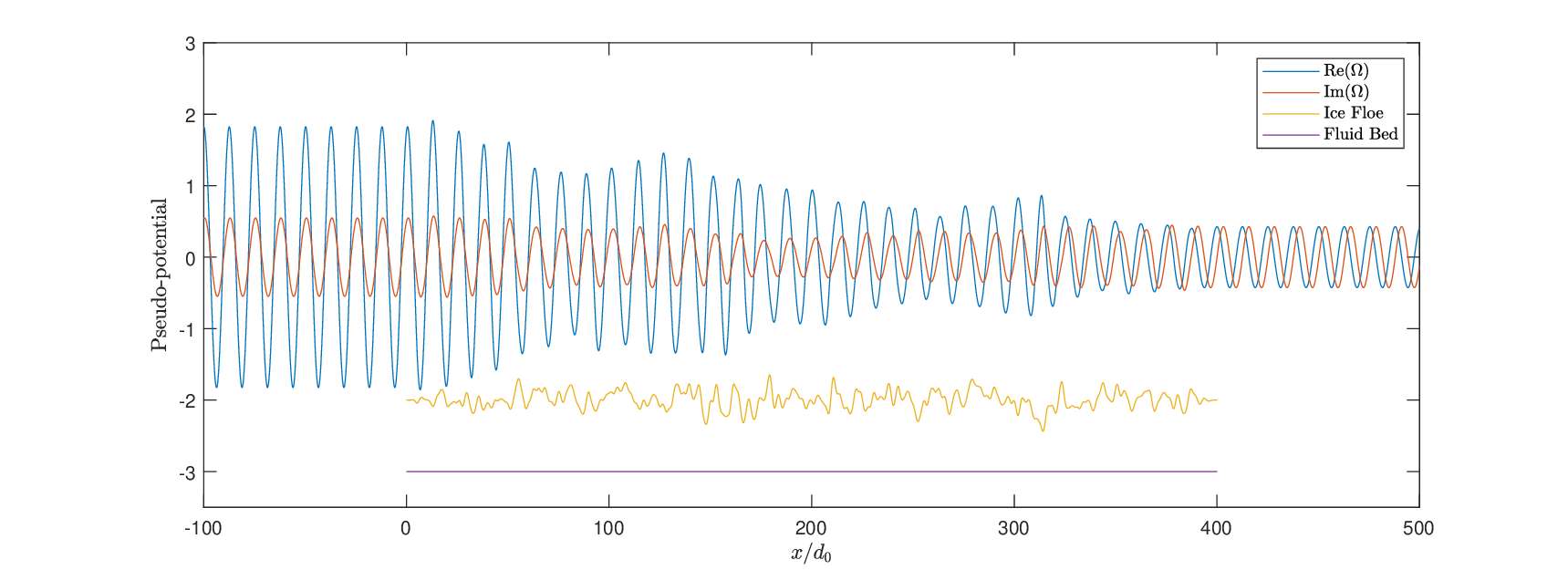}
		\caption{An example of the pseudo-potential (real and imaginary parts of $\Omega(x)$) and an overlay of the random function representing ice submergence across $0 < x < L$. Here, $\sigma^2 = 0.02$, $\Lambda=2d_0$ and $L=400d_0$ and the fluid depth is $h_0 = 2 d_0$.}
		\label{fig8}
	\end{figure}
	
	Having presented theory and simulations in the case of variable bathymetry with no ice cover, we now consider a similar analysis
	of results for a fluid of constant depth $h_0$ covered with floating broken ice submerged to a variable depth $d(x)$, $0 < x < L$, 
	varying randomly about $d_0$, with constant submergence found in 
	$x < 0$ and $x > L$. The only changes from the previous results
	come from different definitions for $C_1$ and $k_0$. Fig.~\ref{fig8} shows the real and
	imaginary parts of the pseudo-potential for a single random 
	simulation of the ice submergence $d(x)/d_0$ illustrated in the
	same plot for which $d_0 = 1$ and $h_0=2d_0$ (the vertical range $(-3,-1)$ is used to represent $(-h_0,0)$.) Again, we observe the signature of partial transmission and reflection in the elevation and note the random response of the pseudo-potential through the variable broken ice cover.
	
	\begin{figure}
		\centering
		\begin{subfigure}[t]{0.03\textwidth}
			\text{(a)}
		\end{subfigure}
		\begin{subfigure}[t]{0.45\textwidth}        \centering
			\includegraphics[width=\linewidth, valign=t]{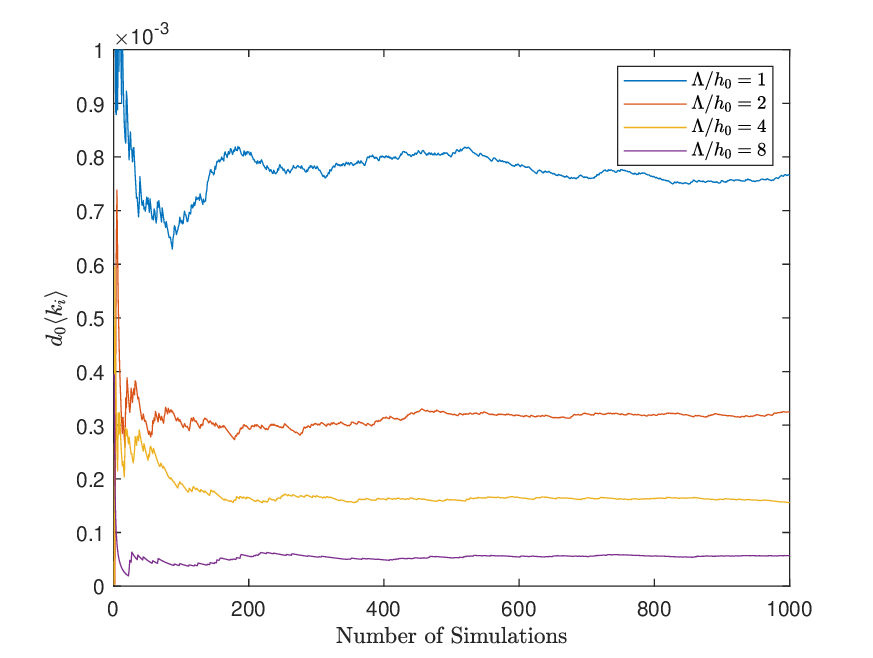}
		\end{subfigure}\hfill
		\begin{subfigure}[t]{0.03\textwidth}
			\text{(b)}
		\end{subfigure}
		\begin{subfigure}[t]{0.45\textwidth}        \centering
			\includegraphics[width=\linewidth, valign=t]{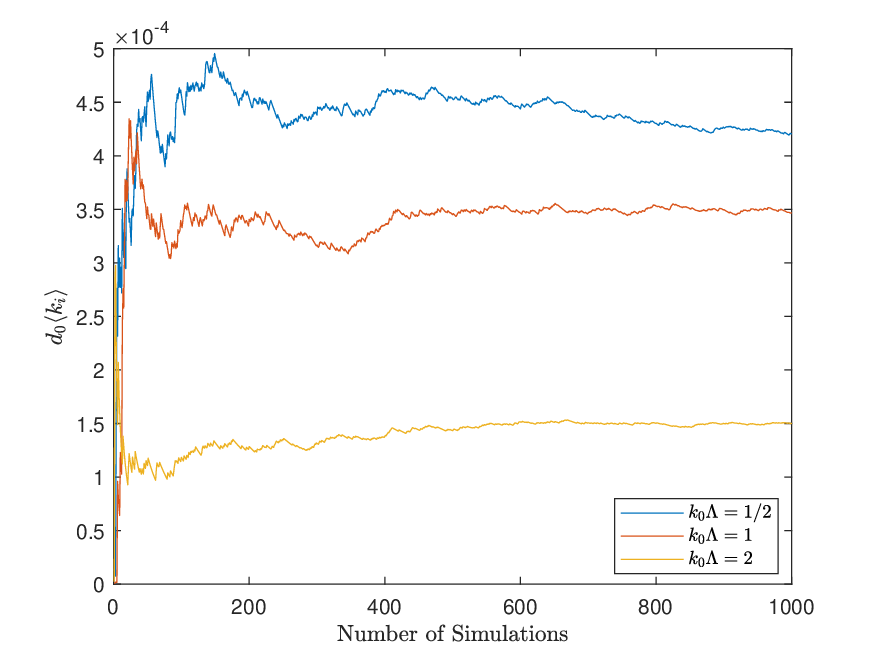}
		\end{subfigure}
		\caption{The variation of the non-dimensional attenuation coefficient with increasing $N$, the number of simulations in the case of randomly varying ice thickness with $\sigma^2 = 0.02$, $L=400d_0$ and $h_0 = 2 d_0$. In (a) $k_0\Lambda=1$ is fixed and $\Lambda/h_0$ is varied; in (b) $\Lambda/h_0 = 4$ is fixed and $k_0 \Lambda$ is varied.}
		\label{fig9}
	\end{figure}
	
	Figure \ref{fig9} illustrates how the ensemble average of the attenuation coefficient converges with $N$, the number of numerical simulations. Each curve is computed from a single set of realisations for particular parameters, but is typical of results across a range of parameters and convergence is identical in character to results for random bathymetry. The depth of the water in these
	and later results, chosen as $h_0 = 2d_0$ may seem small for a physical setting. The primary role of the depth is in setting the wavenumber $k_0$ in terms of the frequency, $K$. The choice $h_0 = 2d_0$ allows us to extend the range of values of $K$ over which the results can be presented without violating the assumptions of shallowness.
	
\begin{figure}
		\centering
		\begin{subfigure}[t]{0.03\textwidth}
			\text{(a)}
		\end{subfigure}
		\begin{subfigure}[t]{0.45\textwidth}        \centering
			\includegraphics[width=\linewidth,valign=t]{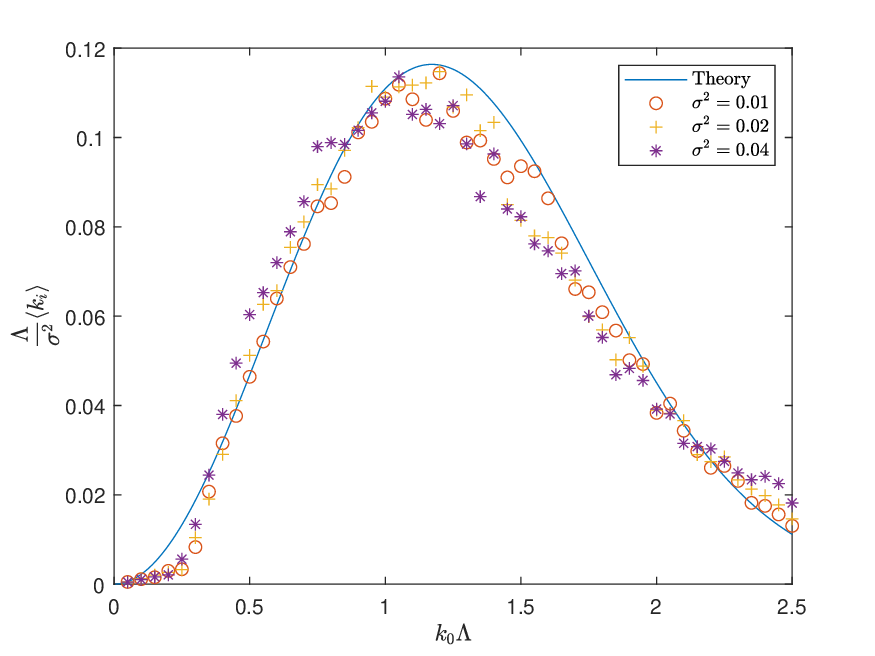}
		\end{subfigure}\hfill
		\begin{subfigure}[t]{0.03\textwidth}
			\text{(b)}
		\end{subfigure}
		\begin{subfigure}[t]{0.45\textwidth}        \centering
			\includegraphics[width=\linewidth,valign=t]{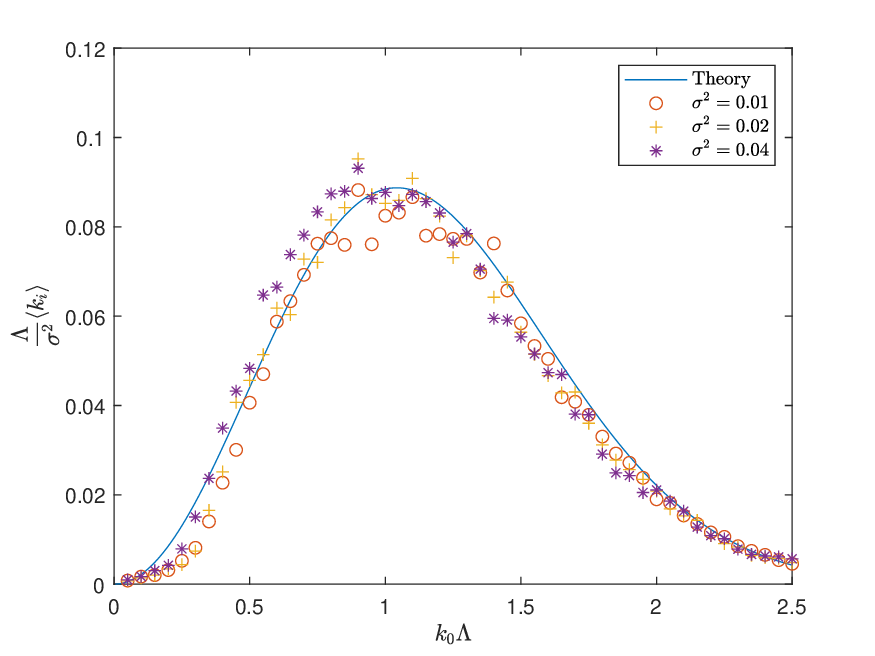}
		\end{subfigure}
		\caption{Scaled attenuation coefficient averaged over $N=500$ simulations of random ice over distance defined by $L = 10 \Lambda/\sigma^2$ compared with theoretical predictions. Here, $h_0 = 2d_0$, $\sigma$ is varied (see legend) and (a) $\Lambda=2d_0$, (b) $\Lambda=4d_0$.}
		\label{fig10}
	\end{figure}

	Figure \ref{fig10} shows results which are analogous to those obtained in Figure \ref{fig5}, comparing the attenuation coefficient calculated by ensemble averaging numerically-determined decay over 500 realisations of a long finite variable ice cover against theoretical results. The vertical axis is scaled so that results for different values of $\sigma$ can be collapsed onto a single theoretical curve. The results for random ice cover differ from those for random
	bathymetry only in the definition of $k_0$ and $C_1$ for ice. 
	
	\subsection{Relationship with other models and field data}
	
	The shallow-water setting and the low-frequency homogenisation used to replace floes of small finite width by a continuum implies that it is inappropriate to make direct comparisons with field data and existing theoretical models especially at very low or high frequencies. However, it is useful to comment on the general features exhibited by our model of wave propagation through broken ice.
	
	The average attenuation coefficient (\ref{eqn3.41}) scales like $k_0^2$ for 
	$k_0 \Lambda \ll 1$ (i.e. at low frequencies) and since 
	$k_0 \propto \omega$ from (\ref{eqn1.2}) in the shallow-water setting, the 
	attenuation scales like $\omega^2$ at low frequencies.
	The attenuation coefficient peaks at $k_0 \Lambda \approx 1$ and then
	decays exponentially for $k_0 \Lambda > 1$. One of the requirements
	of homogenisation is that $d(x)$ varies sufficiently slowly and
	not significantly faster than the wavelength. This translates into
	the condition $k_0 \Lambda \not\ll 1$ and so the peak in
	the attenuation is consistent with the assumptions of the model.
	
	Ongoing work which extends the shallow-water theory presented here to
	deep water (but retaining a homogenisation of the ice floe cover)
	results in attenuation which is proportional to
	$\omega^8$ at low frequencies, whilst a peak and a 
	high-frequency exponential tail remains.
	
	There has been longstanding interest (see, for example, \citet{squire_etal_1995}) in developing a plausible model which captures the relationship 
	between wave frequency and attenuation observed in field measurements.
	Analysis of historical data by \citet{Meylan_Bennetts_Mosig_Rogers_Doble_Peter_2018} suggest attenuation scales like $\omega^n$ for $n$ between 2 and 4 with variations away from this at high and low frequencies. A simple power-law relationship across all frequencies and all ice conditions may therefore not be 
	appropriate. Attenuation of wave energy as it propagates over shallow water or through broken ice is contributed to by both multiple-scattering induced localisation and natural physical dissipation. The primary driver of attenuation in broken ice is unclear (see, e.g. \citet{BENNETTS20121} and \citet{MEYLAN2021101779}) and this paper has only attempted to evaluate multiple-scattering effects. Previous attempts at modelling of attenuation based on 
	multiple scattering through variations in ice thickness (see, e.g. Fig. 4 of 
	\citet{squire2009} and \citet{MEYLAN2021101779}) suggest that, at very low frequencies, the attenuation may scale like
	$\omega^n$ where $n$ is between 8 and 10. However, neither these
	studies nor other multiple-scattering models (see the discussion
	in the Introduction)
	have captured a peak and ``rollover effect'' in the attenuation
	at higher frequencies as we have done in our theory. This may be
	because the onset of rollover occurs at frequencies beyond the limitations
	of our theory. It may also be due to differences in how the 
	multiple-scattering calculations are made in this work compared
	to others. Beyond the assumptions of homogenisation we make no
	other approximation to the scattering process within the ice. 
	In the work of \citet{squire2009}, etc, it is typical 
	that the ice is modelled as a thin elastic plate with no draft and 
	that scattering is calculated using a serial approximation 
	(see \citet{WilliamsThesis}) which effectively neglects reflections at 
	ice floe interfaces and is based on wide-spacing approximations.
	In particular, this latter assumption formally requires breaks 
	in the ice to be large compared to the wavelength and is 
	complementary to our assumption.
	
	The rollover effect that appears in our theory of
	random multiple scattering has been a feature of many
	sets of field measurements taken in the Marginal Ice Zones. 
	See, for example, \citet{squire_etal_1995}, 
	who include field measurements of \citet{wadhams_etal_1988} and \citet{liu_etal_1992} in which attenuation is observed to peak and start to drop as the 
	frequency increases beyond a critical value. High-frequency rollover 
	effects have since been disputed, most notably in 
	\citet{rogers_etal_2016} and \citet{thomson_etal21} who attributes 
	rollover to a statistical effect in data analysis. Thus 
	\citet{thomson_etal21} consider a synthetic (not floating ice) problem 
	in which the attenuation is known and show that measurements fail to replicate
	the expected high-frequency behaviour and, instead, exhibit a rollover effect.

\section{Conclusions}\label{Conclusions}

The paper has considered a basic model for the propagation of long waves through water of variable shallow depth with a surface covered by fragmented broken ice. Simple expressions have been derived for the attenuation of waves over randomly-varying bathymetry and through ice of randomly-varying thickness. In the analytic derivation of the expression for attenuation based on randomness occupying a semi-infinite domain, we have identified and removed terms responsible for incoherent phase cancellations in the ensemble averaging process which contribute to fictitious decay not experienced by individual realisations of wave propagation through randomness. The theory has been shown to agree with numerical simulations in which averaging was performed over individual wave realisations across randomness of finite extent. In the simulations, for which our shallow-water models require numerical solutions to simple two-dimensional ODEs, attenuation was measured accurately by computing eigenvalues of the resulting transfer matrix. These encode propagation but exclude multiple scattering effects relating to transitions at the ends of the scattering region from variable to constant parameter values.

In addition to resolving the discrepancy between theory and numerical simulations for random bathymetry highlighted by \citet{Bennetts_Peter_Chung_2014}, we have also shown that there is a peak in attenuation which relates closely to a Bragg resonant effect, the significant lengthscale of the bed being its statistical correlation length. Beyond this peak, attenuation decreases exponentially as a function of the square of the wavenumber. This decay, predicted by the shallow-water model, therefore appears not to be a finite-depth effect as proposed in some previous studies (e.g. \citet{Devillard_Dunlop_Souillard_1988}, \citet[Section~5]{Mei_Stiassnie_P._2005}).

The shallow-water formulation has been extended to include the effect of broken ice using the method of \citet{porter_2019}. This second-order extension of the classical shallow-water model includes vertical acceleration which is needed for the ice thickness to enter the dynamics. Agreement has been confirmed between theory and numerical simulations. 


Whilst our model may not be applicable to field data due to it being highly simplified, it does provide evidence for a key, albeit disputed, feature of the data sets in that of a ``rollover effect''. This gives us reason to believe that random variations in ice thickness could be a plausible mechanism for the attenuation of waves through broken sea ice, however as the sea ice is multi-phase and non-continuous and our model is limited to a shallow-water model in a continuum ice cover limit further work is needed to establish greater certainty. We plan a range of extensions to the current work to include more complex effects which include: (i) finite water depth; (ii) variable ice cover concentration; (iii) discrete ice floe models; (iv) weak non-linearity and (v) three-dimensional scattering.

\textbf{[Funding]} {L.D. is grateful for the support of an EPSRC (UK) studentship.}\\

\textbf{[Declaration of interests]} {The authors report no conflict of interest.}\\

\textbf{[Data availability statement]} {The data used to produce the figures in this study are openly available at \url{https://doi.org/10.6084/m9.figshare.23912769.v2}}\\

\textbf{[Author ORCIDs]}\\

	\orcidlink{0000-0003-1009-0946} L. Dafydd, https://orcid.org/0000-0003-1009-0946\\
	\orcidlink{0000-0003-2669-0188} R. Porter, https://orcid.org/0000-0003-2669-0188

\section*{Appendix: Derivation of the long wave model}\label{Appendix}
\setcounter{equation}{0}
\renewcommand{\theequation}{A.\arabic{equation}}

The model will be developed in a two-dimensional Cartesian framework $(x,z)$ with $z$ directed vertically upwards. Fluid of density $\rho$ is bounded below by a rigid bed located at $z=-h(x)$ and above by freely-floating fragmented ice of thickness $d(x) \rho/\rho_i$ where $\rho_i$ is the density of ice. The moving fluid/ice interface is described by $z = -d(x) + \zeta(x,t)$ where $\zeta(x,t)$ represent the wave elevation and $t$ is time. Thus the rest position of an unloaded fluid surface would be $z=0$. 

We assume that the depth is small compared to the wavelength and that gradients of $h(x)$ and $d(x)$ are equally small. The ice is assumed broken into individual floes whose horizontal extent is small compared to the wavelength. The floes are constrained to move vertically. The length of individual floes does not enter our model since we assume a continuum model from the outset (the description of the ice submergence as $d(x)$ already indicates this) which avoids engaging in a formal derivation based on multiple horizontal scales.


The fluid is assumed to be both inviscid and incompressible and its motion is represented by the velocity field $(u(x,z,t),w(x,z,t))$, $u$ and $w$ being the horizontal and vertical components of the flow respectively.

Within the fluid, conservation of mass requires
\begin{equation}
u_x + w_z = 0
\label{eqnA.1}
\end{equation}
is satisfied. Conservation of momentum gives
\begin{equation}
\rho u_t + \rho (u u_x + w u_z) = -p_x, \qquad \mbox{and} 
\qquad \rho w_t + \rho (u w_x + w w_z) = -p_z
\label{eqnA.2}
\end{equation}
where $p(x,z,t)$ is the {\it dynamic} pressure in the fluid in excess of background hydrostatic pressure $-\rho g z$ where $g$ is acceleration due to gravity and the 
background atmospheric pressure above the ice is assumed without loss of generality to be zero. On the rigid bed, the no flow condition is represented by
\begin{equation}
w + h'(x) u = 0, \qquad \mbox{on $z = -h(x)$},
\label{eqnA.3}
\end{equation}
and on the moving fluid/ice interface we have the kinematic and dynamic conditions
\begin{equation}
\zeta_t = w +  (d'(x)-\zeta_x(x,t)) u, \qquad \mbox{on $z = -d(x) + \zeta(x,t)$},
\label{eqnA.4}
\end{equation}
and
\begin{equation}
\rho d(x) \zeta_{tt} = p(x,-d(x)+\zeta(x,t),t) - \rho g \zeta(x,t).
\label{eqnA.5}
\end{equation}

We rescale physical variables using
\begin{equation}
x = Lx^*, \quad z = Hz^*,\quad h=Hh^*, \quad d=Hd^*\quad \mbox{and}\quad \zeta = A\zeta^*,
\label{eqnA.6}
\end{equation}
where $L$ represents a characteristic horizontal lengthscale (a different definition from the one used in the main part of the text for the length of the bed) associated with the wavelength and/or the variable bed/ice cover, $H$ is a characteristic fluid depth and $A$ a characteristic wave elevation. We also define
\begin{equation}
\epsilon = \frac{H}{L},\qquad \delta=\frac{A}{H}
\label{eqnA.7}
\end{equation}
which represents shallowness and 
linearisation parameters respectively. We suppose that both $\epsilon$ and $\delta$ are small and assume that $\delta=o(\epsilon^2)$ to ensure we operate within a 
linearised setting.

Based on the shallow water dispersion relation, we select a timescale $T=L/\sqrt{gH}$ so that $t= Lt^*/\sqrt{gH}$ and set
\begin{equation}
u = \frac{A}{H}\sqrt{gH}u^* \qquad \mbox{and} \qquad w = \frac{A}{L}\sqrt{gH}w^*
\label{eqnA.8}
\end{equation}
whilst $p = \rho g A p^*$. Under this change of variables the governing equations become
(after dropping asterisks)
\begin{equation}
u_{x} + w_{z} = 0
\label{eqnA.9}
\end{equation}
with
\begin{equation}
u_{t} + \delta (u u_{x}+w u_{z}) = -p_{x}
\label{eqnA.10}
\end{equation}
and
\begin{equation}
\epsilon^2 w_{t} + \delta \epsilon^2 (u w_{x}+w w_{z}) = -p_{z}.
\label{eqnA.11}
\end{equation}
Our boundary condition at the fluid bed reads
\begin{equation}
w + h'(x) u = 0 \qquad \mbox{on $z=-h(x)$}
\label{eqnA.12}
\end{equation}
with our boundary conditions on the ice becoming
\begin{equation}
\zeta_{t} = w + (d'(x) - \delta\zeta_x(x,t)) u, \qquad \mbox{on $z = -d(x) + \delta \zeta(x,t)$}
\label{eqnA.13}
\end{equation}
and
\begin{equation}
\epsilon^2 d(x) \zeta_{tt} = p(x,-d(x) + \delta\zeta(x,t),t)- \zeta.
\label{eqnA.14}
\end{equation}
Noting that $\delta = o(\epsilon^2)$ has been assumed we expand variables up to $O(\epsilon^2)$, so that
\begin{equation}
\zeta(x,t) = \zeta^{(0)}(x,t) +\epsilon^2\zeta^{(1)}(x,t) +\ldots
\label{eqnA.15}
\end{equation}
and
\begin{equation}
\{ p,u,w\}(x,z,t) = \{p^{(0)}, u^{(0)}, w^{(0)} \}(x,z,t) 
+\epsilon^2 \{ p^{(1)}, u^{(1)}, w^{(1)} \}(x,z,t) + \ldots.
\label{eqnA.16}
\end{equation}
Only in the case that $h(x)$ and/or $d(x)$ contain discontinuities would we need
to include terms of $O(\epsilon)$ (see, \citet[Section~5]{Mei_Stiassnie_P._2005}) since these would arise from
an asymptotic matching process across the discontinuity.
It is consistent with this expansion that we neglect contributions from terms multiplying $\delta$ in (\ref{eqnA.9})-(\ref{eqnA.14}).
We continue by solving for the leading order variables. From (\ref{eqnA.11}), $p^{(0)}_z = 0$ and from (\ref{eqnA.14}), $p^{(0)}(x,-d(x),t) = \zeta^{(0)}(x,t)$ implies
\begin{equation}
p^{(0)}(x,z,t) = \zeta^{(0)}(x,t)
\label{eqnA.17}
\end{equation}
and then from (\ref{eqnA.10}) we have
\begin{equation}
u_t^{(0)}(x,z,t) = -\zeta^{(0)}_x(x,t)
\label{eqnA.18}
\end{equation}
and so $u^{(0)}$ is a function of $x$ and $t$ only.
Integrating (\ref{eqnA.9}) at leading order from $z=-h(x)$ to $z=-d(x)$ and using 
(\ref{eqnA.12}) and (\ref{eqnA.13}) gives
\begin{equation}
q^{(0)}_x(x,t) = \left( (h(x)-d(x))u^{(0)}(x,t) \right)_x = -\zeta_t^{(0)}(x,t)
\label{eqnA.19}
\end{equation}
where we have defined the depth-integrated horizontal fluid flux $q(x,t) = q^{(0)}(x,t) + \epsilon^2 q^{(1)}(x,t) + \ldots$ with
\begin{equation}
q^{(0,1)}(x,t) = \int_{-h(x)}^{-d(x)} u^{(0,1)} (x,z,t) \, dz.
\label{eqnA.20}
\end{equation}
Eliminating between (\ref{eqnA.18}) and (\ref{eqnA.19}) gives either
\begin{equation}
\zeta_{tt}^{(0)} = \left( (h(x)-d(x)) \zeta^{(0)}_x \right)_x,
\qquad \mbox{or} \qquad
q_{tt}^{(0)} = (h(x)-d(x)) q_{xx}^{(0)}
\label{eqnA.21}
\end{equation}
as the leading order governing equation, expressed in dimensionless variables. That is, the effect of fragmented ice cover at leading order is equivalent to an uncovered fluid having a reduced depth, $h(x)-d(x)$.

Now we work at the next order, $O(\epsilon^2)$. Integrating (\ref{eqnA.9}) at order $O(\epsilon^2)$ from $z=-h(x)$ to $z=-d(x)$ and using (\ref{eqnA.12}) and (\ref{eqnA.13}) 
at $O(\epsilon^2)$ gives
\begin{equation}
q^{(1)}_x(x,t) = \frac{\partial}{\partial x} \int_{-h(x)}^{-d(x)} u^{(1)}(x,z,t) \, dx = -\zeta_t^{(1)}(x,t).
\label{eqnA.22}
\end{equation}
The next step is to determine the leading order vertical velocity integrating (\ref{eqnA.9}) again, but now from $z$ to $-d(x)$ to give
\begin{equation}
w^{(0)}(x,z,t) = \zeta^{(0)}_t(x,t) - \left( (z+d(x)) u^{(0)}(x,t) \right)_x
\label{eqnA.23}
\end{equation}
which is linear in $z$. From (\ref{eqnA.11}) at $O(\epsilon^2)$ we infer that
\begin{equation}
p^{(1)}_z(x,z,t) = -\zeta_{tt}^{(0)} + \left( (z+d(x)) u_t^{(0)} \right)_x
\label{eqnA.24}
\end{equation}
which can be integrated using the condition (\ref{eqnA.14}) at $O(\epsilon^2)$ to give
\begin{equation}
p^{(1)}(x,z,t) = \zeta^{(1)} - z \zeta_{tt}^{(0)} + \frac12 \left( (z+d(x))^2 u_t^{(0)} \right)_x.
\label{eqnA.25}
\end{equation}
Using in (\ref{eqnA.10}) at $O(\epsilon^2)$ gives
\begin{equation}
u_t^{(1)}(x,z,t) = -p_x^{(1)} = z \zeta_{ttx}^{(0)} -\zeta^{(1)}_x - \frac12 \left( (z+d(x))^2 u_t^{(0)} \right)_{xx}.
\label{eqnA.26}
\end{equation}
We find, after extensive algebra, which makes repeated use of the relation $q^{(0)}_t = (h-d) u^{(0)}_t$, that
\begin{multline}
q_t^{(1)}(x,t) = \int_{-h(x)}^{-d(x)} u^{(1)}_t \, dz = 
\frac12 (d^2-h^2) \zeta_{ttx}^{(0)} - (h-d) \zeta_x^{(1)}
+ \frac12 (h-d) d'' q_t^{(0)}
\\
-\frac16 \left\{ (h-d)' q_{xxt}^{(0)} -2 (h-d)(h'-d') q_{xt}^{(0)} -
(h''-d'') (h-d) q_t^{(0)} +2 (h'-d')^2 q_t^{(0)} \right\}
\\ 
- {d'}^2 q_t^{(0)} + d' \left\{ (h-d)q_{xt}^{(0)} - (h' - d') q_t^{(0)}  \right\}.
\label{eqnA.27}
\end{multline}
Further simplification and use of the relation $q_x^{(0)} = -\zeta_t^{(0)}$ results in 
\begin{multline}
q_t^{(1)} = - (h-d) \left( \zeta^{(1)}_x + \frac13 \left( (h + 2d) \zeta_{tt}^{(0)} \right)_x
\right) 
\\ 
+ q_t^{(0)} \left( \frac16 (h-d) (h + 2 d)'' - \frac13 (h-d)'(h+2d)' -d'^2
\right).
\label{eqnA.28}
\end{multline}
We can now recombine leading order and $O(\epsilon^2)$ terms as we redimensionalise variables, a process which leads to the coupled equations
\begin{equation}
\zeta_t = - q_x
\label{eqnA.29}
\end{equation}
and
\begin{equation}
\left( 1 +d'^2 + \frac13 (h-d)'(h+2d)' - \frac16 (h-d) (h + 2 d)'' \right) q_t
= -(h-d) \left( g \zeta + \frac{(h+2d)}{3} \zeta_{tt} \right)_x
\label{eqnA.30}
\end{equation}
expressed in terms of the original physical variables $q$ and $\zeta$ and which are accurate to $O(\epsilon^2)$. Eliminating $q$ in favour of $\zeta$ gives us the governing equation
\begin{equation}
\zeta_{tt} = \frac{\partial}{\partial x} \left( \hat{d}(x) \frac{\partial}{\partial x}
\left( g \zeta + \frac{(h+2d)}{3} \zeta_{tt} \right) \right)
\label{eqnA.31}
\end{equation}
where
\begin{equation}
\hat{d}(x) = \frac{(h-d)}{1 +d'^2 - \frac16 (h-d)(h+2d)'' + \frac13 (h-d)'(h+2d)'}.
\label{eqnA.32}
\end{equation}
Note that when $d(x) \equiv 0$ we recover equation (2.13) from \citet{porter_2019}. We see that
the expansion to $O(\epsilon^2)$ in the small parameter $\epsilon = H/L$ has captured the
contribution from the inertia of the ice in (\ref{eqnA.31}) whilst there are non-trivial modifications to the wave speed through the geometrical factors associated with varying $d(x)$ and $h(x)$ in (\ref{eqnA.32}).
Specifically, it is worth noting that $(h+2d)/3 = (h-d)/3 + d$ and $h-d$ is the vertical 
extent of the fluid. Thus, the isolated contribution $d \zeta_{tt}$ is associated with 
ice inertia and the remaining $\frac13 (h-d) \zeta_{tt}$ is a contribution from vertical acceleration of the fluid through depth-averaging, in common with \citet{porter_2019}.

Eliminating $\zeta$ in favour of $q$ between (\ref{eqnA.29}) and (\ref{eqnA.30}) gives
\begin{equation}
q_{tt} = \hat{d}(x) \left( g q_x + \frac{(h+2d)}{3} q_{ttx} \right)_x
\label{eqnA.33}
\end{equation}
and this provides the starting point for a series of transformations of the dependent
variable which follow \citet{porter_2019}. We factorise a time-harmonic variation with 
\begin{equation}
q(x,t) = \Re \left\{ \frac{\varphi(x)}{\sqrt{1- \frac13 K (h + 2d)}} 
\exp^{-\ci \omega t} \right\}
\label{eqnA.34}
\end{equation}
and the square-root factor in the denominator simultaneously transforms the resulting ODE into canonical form. Thus, after some algebra we find
\begin{equation}
\varphi''(x)+\left(\frac{\hat{K}}{h-d}\left(1+\frac13  v_1(h,d) h'(x)^2 + \frac13 v_2(h,d) (d'(x)^2 + h'(x)d'(x))\right)\right)\varphi(x) = 0
\label{eqnA.35}
\end{equation}
where
\begin{equation}
\hat{K} = \frac{K}{1 - \frac13 K (h + 2d)},
\label{eqnA.36}
\end{equation}
\begin{equation} v_1(h,d) = 1 + \frac{1}{12}\hat{K}(h(x)-d(x))
\qquad \mbox{and}
\qquad v_2(h,d) = 1 + \frac{1}{3}\hat{K}(h(x)-d(x)).
\label{eqnA.37}
\end{equation}

A final change of variables is made, by letting $\Omega(x)=\varphi'(x)$ and it follows that (\ref{eqnA.35}) is transformed into
\begin{equation}
(\doublehat{d}(x) \Omega')' + K \Omega = 0
\label{eqnA.38}
\end{equation}
where
\begin{equation}
\doublehat{d}(x) = \frac{(h-d)(1-\frac{1}{3}K(h+2d))}{1+\frac13  v_1(h,d) h'(x)^2 + \frac13 v_2(h,d) (d'(x)^2 + h'(x)d'(x))}.
\label{eqnA.39}
\end{equation}
This final series of transformations have brought about two useful features. The first is that (\ref{eqnA.38}) is expressed in a form aligned with the familiar linearised first order shallow water equation. The second is that the function $\Omega(x)$ and its derivative $\Omega'(x)$ are continuous even if $h'(x)$ and/or $d'(x)$ are discontinuous. 
The loaded surface can be reconstructed from $\Omega(x)$ by following the 
effect of each transformation and turns out to be represented by
\begin{equation}
\label{eqnA.40}
\eta =\frac{(-i/\omega)}{\sqrt{1-\frac{1}{3}K(h+2d)}}\left(\Omega(x)-\frac{\frac{1}{6}(h-d)(h+2d)'}{1+\frac13  v_1(h,d) h'(x)^2 + \frac13 v_2(h,d) (d'(x)^2 + h'(x)d'(x))}\Omega'(x)\right)
\end{equation}
where $\zeta(x,t) = \Re \{ \eta(x) \exp^{-\ci \omega t} \}$.

Since we anticipate $K h \ll 1$, we can make approximations $v_1(h,d) \approx 1$ and $v_2(h,d) \approx 1$, noting $0<h-d\leq h$ and so
\begin{equation}
\frac13  v_1(h,d) h'(x)^2 + \frac13 v_2(h,d) (d'(x)^2 + h'(x)d'(x))\approx \frac13 (h'(x)^2+h'(x)d'(x) + d'(x)^2).
\label{eqnA.41}
\end{equation}
We note that if we let $d(x)=0$ in (\ref{eqnA.39}), (\ref{eqnA.40}) and (\ref{eqnA.41}) we recover expressions derived in \citet{porter_2019}.

\bibliographystyle{jfm}
\bibliography{bibliography}

\end{document}